\documentclass[
  aps,
  prl,
  twocolumn,
  superscriptaddress,
  longbibliography
]{revtex4-2}

\usepackage[whole]{bxcjkjatype}

\usepackage{amsmath,amssymb,mathtools,bm,braket}
\usepackage{graphicx}
\usepackage{xcolor}
\usepackage{comment}
\usepackage[colorlinks=true,linkcolor=blue,citecolor=blue,urlcolor=blue]{hyperref}
\usepackage{mathrsfs}

\newcommand{\Tr}{\operatorname{Tr}}

\newcommand{\diam}{\operatorname{diam}}
\newcommand{\dist}{\operatorname{dist}}
\newcommand{\id}{\mathbf{1}}
\newcommand{\can}{\mathrm{ens}}
\newcommand{\MITE}{\mathrm{MITE}}
\newcommand{\QC}{\mathrm{QC}}
\newcommand{\opt}{\mathrm{opt}}
\newcommand{\loc}{\mathrm{loc}}
\newcommand{\ee}{\mathrm{e}}
\newcommand{\ii}{\mathrm{i}}
\newcommand{\Torder}{\mathcal{T}}
\newcommand{\PRLsection}[1]{\par\medskip\noindent\textit{#1.—}\ }

\usepackage{tcolorbox}
\tcbuselibrary{theorems,breakable}

\newtcbtheorem[number within=section]{theo}{Theorem}%
{breakable,colback=green!5,colframe=green!35!black,fonttitle=\bfseries}{th}

\newtcbtheorem[number within=section]{defi}{Definition}%
{colback=blue!5, colframe=blue!35!black, fonttitle=\bfseries}{def}

\begin{document}

\title{Information Thermodynamics in Closed Quantum Many-Body Dynamics}

\author{Yusuke O. Nakai}
\affiliation{RIKEN Center for Emergent Matter Science (CEMS), RIKEN,
Wako, Saitama 351-0198, Japan }
\affiliation{Nonequilibrium Quantum Statistical Mechanics RIKEN Hakubi Research Team, RIKEN Pioneering Research Institute (PRI), 2-1 Hirosawa, Wako, Saitama 351-0198, Japan}

\author{Shoki Sugimoto}
\affiliation{Department of Applied Physics, The University of Tokyo, 7-3-1 Hongo, Bunkyo-ku, Tokyo 113-8656, Japan}
\affiliation{Nonequilibrium Quantum Statistical Mechanics RIKEN Hakubi Research Team, RIKEN Pioneering Research Institute (PRI), 2-1 Hirosawa, Wako, Saitama 351-0198, Japan}

\author{Ryusuke Hamazaki}
\affiliation{Nonequilibrium Quantum Statistical Mechanics RIKEN Hakubi Research Team, RIKEN Pioneering Research Institute (PRI), 2-1 Hirosawa, Wako, Saitama 351-0198, Japan} 
\affiliation{RIKEN Center for Interdisciplinary Theoretical and Mathematical Sciences (iTHEMS), 2-1 Hirosawa, Wako, Saitama 351-0198, Japan}

\date{\today}

\begin{abstract}
Although information thermodynamics explains how the second law is modified due to the information acquired by Maxwell's demon, its formulation in many-body systems remains largely open. Here, we develop an information-thermodynamic framework for closed quantum many-body systems initially in microscopic thermal equilibrium (MITE), under physically motivated restrictions: local quantum measurements and finite-time controls generated by local Hamiltonians. 
By additionally assuming that each local feedback term depends only on nearby measurement outcomes,
we rigorously prove that the extractable work coincides in the thermodynamic limit with that extractable from the corresponding canonical ensemble.
We further express this extractable work in terms of quantum-information quantities, including the quantum-classical mutual information and the Holevo information for the corresponding canonical ensemble. 
We also show that feedback based on distant measurement outcomes can exploit long-range correlations hidden in MITE, yielding an extensive increase in extracted work relative to the canonical ensemble.
\end{abstract}

\maketitle

\PRLsection{Introduction}
Maxwell's demon is a thought experiment in which feedback control based on measurement outcomes appears to violate the second law of thermodynamics \cite{maxwell1871,bennett1982,szilard1929,landauer1961}. 
In modern information thermodynamics, the paradox is resolved by incorporating the information acquired through measurement and the cost of erasing memory~\cite{sagawa-ueda2008PRL,Sagawa2009erasure,sagawa2010Jarzynski,sagawa2012second}, and
the last decades have witnessed remarkable success in this field especially for few-body systems amenable to precise control~\cite{Kim2011,Parrondo2015,Elouard2017,Manzano2018,Toyabe2010,Koski2014,Masuyama2018}.
However, despite recent progress~\cite{YYS2025,YYS2026}, many questions remain as to how information thermodynamics should be formulated in many-body systems, the demon's original setting. 
Such questions have especially regained importance in light of recent experimental advances in quantum computers and quantum simulators, which have made measurement-based feedback control of otherwise unitary many-body dynamics feasible \cite{Riste2013,Daniel2016,Manuel2016,Negnevitsky2018}.
It is therefore important to explore how information thermodynamics extends to closed quantum many-body systems.

In parallel, the extractable work in
closed quantum dynamics (without measurement and feedback)
has recently been studied extensively as a fundamental problem in quantum thermodynamics and quantum-battery research ~\cite{allahverdyan-etal2004,alicki-fannes2013,Skrzypczyk2014,Binder2015,Perarnau2015,Campaioli2017,Manzano2018,Andolina2019,Francica2020,Touil2022,Mitsuhashi2022,campaioli-etal2024,Hokkyo2025,Medina2025,George2025,sugimoto2026,Watanabe2026,Somnath2026clifford}.
The maximum work extractable under arbitrary unitary operations is called ergotropy \cite{allahverdyan-etal2004}, and a state whose energy expectation value cannot be lowered by any unitary operation is called a passive state~\cite{pusz-woronowicz1978,lenard1978}.
This property is an (idealized) formulation of Planck's principle, which provides one way to characterize quantum thermal equilibrium.
The canonical ensemble $\rho^\can=\ee^{-\beta H}/Z$ with $\beta\geq 0$ is passive, and no work can be extracted from it by unitary operations alone.

In a closed quantum many-body system, however, not only a canonical ensemble but also pure states can represent thermal equilibrium with respect to local observables.
One of such classes of states is microscopic thermal equilibrium (MITE), where the expectation value of every local observable agrees with that in the corresponding canonical ensemble \cite{MITE}.
Importantly, such thermal equilibrium with respect to local observables does not immediately imply passivity in the conventional sense.
Indeed, if arbitrary unitary operations are allowed, we can transform the initial thermal pure state to the ground state, which leads to extensive ergotropy~\cite{allahverdyan-etal2004}.
In realistic quantum many-body systems, however, arbitrary global unitaries cannot be implemented freely. It has recently been shown that no extensive work can be extracted from finite-time operations under local Hamiltonians, i.e., recovering the second law of thermodynamics for thermal pure states~\cite{Hokkyo2025,Chiba2026}.

Against this background, one crucial question is how the second law on work extraction can be modified  under realistic (e.g., local) measurement and feedback operations in closed quantum many-body systems.
We note that related notions, such as daemonic ergotropy, characterize the enhancement of ergotropy by measurement and feedback \cite{Francica2017Daemonic}.
However, it remains unclear how such measurement-assisted work extraction should be formulated thermodynamically in closed many-body systems (using, e.g., temperature), especially for thermal pure states under locality constraints.

In this Letter, we develop an information-thermodynamic framework
for closed quantum many-body systems initially in MITE under
experimentally and physically motivated restrictions: local quantum
measurements and finite-time unitary controls generated by local Hamiltonians.
When we additionally assume each local term in the feedback Hamiltonian to depend only on measurement outcomes obtained in nearby regions, 
we rigorously prove that
the work extractable from MITE (a pure state) coincides in the thermodynamic limit with that extractable from the canonical ensemble (a mixed state).
The latter is expressed in terms of the QC mutual information
\cite{QC1,QC2,QC3,sagawa-ueda2008PRL}, the Holevo information
\cite{holevo1,holevo2,holevo3}, and non-equilibrium free energy.
Thus, even when the entire system is in a pure state from which macroscopic work could be extracted under unrestricted operations, the work extractable under these constraints is bounded in terms of measurement-related information quantities.
In contrast, if we allow feedback based on
distant measurement outcomes, we show that long-range correlations hidden in a pure MITE state
can be exploited for work extraction exceeding that from the canonical ensemble despite finite-time quantum operations.
Importantly, our results are relevant to quantum simulators and
quantum computers, where pure many-body states are naturally prepared
whereas preparing an exact canonical ensemble is generally challenging.

\PRLsection{Measurement-and-feedback protocol in closed many-body systems}
We consider the work extractable by measurement and feedback control applied to a closed quantum many-body system of volume $V=L^d$.
Here, a closed quantum many-body system exchanges no heat with a heat bath but is subject to external operations consisting of measurement and feedback.
We focus on a cyclic protocol, and take the initial and final Hamiltonians to be
$
 H_i=H_f=H=\sum_X h_X,$
where $h_X$ is a local operator supported on a subregion $X$, whose diameter ${\rm diam}(X)$ does not depend on the system size $L$.
Here, we define the diameter of the region $X$ by
$
    {\rm diam}(X):=\max\{{\rm dist}(x,y):x,y\in X\},
$
where ${\rm dist}(x,y)$ is the Euclidean distance.

We consider a protocol similar to that in~\cite{sagawa-ueda2008PRL}, where the measurement and feedback are applied only once.
It consists of three steps. First, a unitary operation $U_1$ is applied before the measurement. Next, a measurement described by Kraus operators $\{M_{\bm k}\}$ is performed, where $\bm k$ is a set of measurement outcomes and $\{M_{\bm k}\}$ satisfies $ \sum_{\bm k}M_{\bm k}^{\dagger}M_{\bm k}=\id$. Finally, conditioned on $\bm k$, a feedback unitary $U_{\bm k}$ is applied. 
For an initial state $\rho_i$, the probability of obtaining outcome $\bm k$ and the postfeedback state conditioned on outcome $\bm k$ are
$ P_{\bm k}
 =\Tr\!\left[M_{\bm k}U_1\rho_iU_1^\dagger M_{\bm k}^\dagger\right]$
 and
 $\rho_{\bm k}'
 =U_{\bm k}M_{\bm k}U_1\rho_iU_1^\dagger
 M_{\bm k}^\dagger U_{\bm k}^\dagger/P_{\bm k}
$,
respectively.
Averaging over the outcomes $\{\bm k\}$ gives the final state
$
 \rho_f=\sum_{\bm k}P_{\bm k}\rho_{\bm k}'
 =\sum_{\bm k}U_{\bm k}M_{\bm k}U_1\rho_iU_1^\dagger
 M_{\bm k}^\dagger U_{\bm k}^\dagger.
$

Because the closed quantum many-body system exchanges no heat with a heat bath, we define the extracted work as the net decrease in the system's energy induced by the sequence of external operations~\cite{sagawa-ueda2008PRL}:
\begin{equation}
 W=\Tr[H\rho_i]-\Tr[H\rho_f].
 \label{eq:work-definition}
\end{equation}
Although a physical implementation of the measurement may exchange energy with the system, the resulting change in the system energy is reflected in $\rho_f$; the energetic costs associated with the measurement apparatus and feedback controller, including information processing and memory reset, are not included in $W$ \cite{sagawa-ueda2008PRL,sagawa2012second}.
A positive value $W>0$ means that energy has been extracted from the system.

Our main assumption is that 
$U_1$ and $U_{\bm k}$ are generated by local Hamiltonians in an
$L$-independent time and that $\{M_{\bm k}\}$ describes local
measurement as specified below.
If we first assume additionally that each local term in the
feedback Hamiltonian depends only on measurement outcomes obtained
within an $L$-independent distance of its support, we obtain our main theorem:
the work $W$ extractable from general initial MITE is given by
\begin{equation}
 \beta W\simeq\beta W_{\can}
 =I_{\QC}-\chi-D(\rho_f^\can\Vert\rho_i^{\can}).
 \label{eq:main-information}
\end{equation}
Here, $W_\mathrm{ens}$ denotes the work extracted by applying the
same protocol to the initial canonical ensemble
$\rho_i^{\can}=e^{-\beta H}/Z$, producing the final
(generally nonequilibrium) state $\rho_f^{\can}$.
The symbol $\simeq$ means that subextensive corrections are
neglected for large $L$, and the right-hand side consists of
information-theoretical quantities defined for this protocol
with the canonical initial state.
In the following, we detail the definition of each quantity,
the conditions for the equality, and their meanings.

\PRLsection{Work extraction from MITE under constraints on locality and operational times}
We first discuss the first equality in Eq.~\eqref{eq:main-information}, which serves as a main achievement in our work.
A microscopic thermal equilibrium (MITE) $\rho_i^{\MITE}$~\cite{MITE} with inverse temperature $\beta$ is defined as a (possibly pure) state where expectation values of local observables $O$ are  indistinguishable from those of the canonical ensemble $\rho_i^{\can}$ with the same $\beta$:
\begin{equation}
  \Tr[\rho_i^{\MITE}O]\simeq \Tr[\rho_i^{\can}O].
  \label{eq:MITE}
\end{equation}
Here, $A\simeq B$ means $|A-B|=o(1)$ and $o(V)$ for intensive and extensive $A$, respectively.
To keep the discussion general, we consider an extended notion of MITE and assume that the local observable $O$ has support $S_{\rm loc}$ whose radius scales sublinearly with the system size $L$:
$
    {\rm diam}[S_{\rm loc}]=\mathcal{O}(L^s).
    $
Note that $s=0$ corresponds to the ordinary MITE~\cite{MITE}, but the MITE with $0<s<1$ can also occur for typical states randomly taken from the microcanonical ensemble~\cite{Goldstein2006,sugimoto2023bounds}.
As shown below,   our results hold for each fixed $s \:(0\leq s<1)$.

Next, we write the unitary operation applied before the measurement as
$
  U_1(t'):=\Torder\exp\!\left[-\ii\int_0^{t'}H_1(u)\,du\right]
$
and the feedback operation applied after the measurement as
$    U_{\bm k}(t):=\Torder\exp\!\left[-\ii\int_0^t H_{\bm k}(u)\,du\right],$
where $\mathcal{T}$ denotes the time ordering.
Crucially, we assume that the Hamiltonians $H_1(u)$ and $H_{\bm{k}}(u)$ are given by the sum of local operators (whose supports do not increase with $L$), and the durations of the two unitary operations satisfy $t,t'=\mathcal{O}(L^s)$.

We also impose locality on the measurement.
Specifically, we represent the measurement operator corresponding to outcome $\bm k$ by a depth-$d_{\mathrm M}$ local measurement circuit:
\begin{align}
 M_{\bm k}&=m_{\bm k_{d_\mathrm{M}}}^{d_{\mathrm M}}m_{\bm k_{d_\mathrm{M}-1}}^{d_{\mathrm M}-1}\cdots m_{\bm k_1}^{1},
 \label{eq:measurement-depth}\\
m_{\bm k_n}^{n}&=\bigotimes_{Y\in\mathcal{F}_n}m_{k_{n,Y}}^{n,Y},
 \label{eq:measurement-layer}
\end{align}
where $\bm{k}=(\bm{k}_1,\ldots,\bm{k}_{d_{\mathrm M}})$ denotes the full record of local measurement outcomes with $\bm{k}_n=(k_{n,Y})_{Y\in\mathcal{F}_n}$ collecting the outcomes in the $n$th measurement layer.
Here, $\mathcal{F}_n$ denotes a set of mutually disjoint local regions $Y$, on which $m_{k_{n,Y}}^{n,Y}$ is supported. 
We assume that a measurement represented in this form is low-depth and local, i.e., $(\mathscr{D}-1)d_{\mathrm M}=\mathcal{O}(L^{s})$, where $    \mathscr{D}=\max_{n}\max_{Y\in\mathcal{F}_n}{\rm diam}(Y)+1$ (see Supplemental Material for details \cite{supplement}).

We further impose locality on the dependence of the feedback
Hamiltonian on measurement outcomes.
Physically, this means that the classical communication is spatially local.
We write $H_{\bm k}(u)=\sum_Z\tilde h_{\bm k^{(Z)},Z}(u)$,
where each term is supported on $Z$ and $\bm k^{(Z)}$ collects the
measurement outcomes on which that term depends.
Our requirement means that
$\bm k^{(Z)}$ is a set of $k_{n,Y}$ satisfying
\begin{equation}
 \dist(Z,Y)\le\ell,\qquad
 \ell=\mathcal{O}(L^s).
 \label{eq:feedback-outcome-locality}
\end{equation}
Here, $\dist(Z,Y):=\min_{z\in Z,y\in Y}\dist(z,y)$.

Under these conditions, apply the same measurement-and-feedback protocol to $\rho_i^{\can}$ and $\rho_i^{\MITE}$. Then we can prove that 
\begin{equation}
 W\simeq W_{\can}
 \label{eq:main-equivalence}
\end{equation}
holds for arbitrary $\beta$ (including $\beta=0$). 
The rigorous formulation of the theorem and its proof are given in Supplemental Material~\cite{supplement}, where we apply the Lieb-Robinson bound~\cite{Lieb1972,Bravyi2006,gong2022bounds} to the Hamiltonian in the Heisenberg picture.
We emphasize that extending the unitary-only setting~\cite{Hokkyo2025,Chiba2026} to measurement and feedback is nontrivial, since even a local measurement can induce a nonlocal state update at the level of individual quantum trajectories. Nevertheless, under the above constraints, the extracted work
agrees with that obtained from the corresponding canonical ensemble.

\PRLsection{Optimized extractable work}
We next consider feedback operations optimized to maximize the
extractable work.
For fixed $U_1$ and measurement operators $\{M_{\bm k}\}$, we
denote the optimal work extractable from the canonical initial
state under arbitrary unitary feedback by $W_{\can}^{\opt}$.
By definition, $W_{\can}\le W_{\can}^{\opt}$ holds.
Equation~\eqref{eq:main-equivalence} holds for every protocol
satisfying the locality and time constraints stated above,
including Eq.~\eqref{eq:feedback-outcome-locality}.
Therefore,
\begin{equation}
 W\simeq W_{\can}\le W_{\can}^{\opt}.
 \label{eq:daemonic-bound}
\end{equation}
Thus, although MITE may be pure and nonpassive, the work
extractable under these constraints does not exceed the optimal work extractable from
the corresponding canonical ensemble with the same $U_1$ and
measurement operators.

Two remarks are in order.
First, this argument requires all of the above constraints.
If longer operation times, nonlocal Hamiltonians, or dependence
on distant measurement outcomes (i.e., nonlocal classical communication) are allowed, the equivalence
in Eq.~\eqref{eq:main-equivalence} is no longer guaranteed.
As discussed later, we indeed demonstrate the last possibility 
leads to additional work gain from long-range correlations in an entangled pure MITE state.
Second, the equality condition in \eqref{eq:daemonic-bound} is achieved for a certain setup, as discussed later.

\PRLsection{Information-thermodynamic identity}
We now  show the second equality in Eq.~\eqref{eq:main-information}:
\begin{equation}
  \beta W_{\rm ens}=I_{\QC}-\chi-D(\rho_f^\can\Vert\rho_i^{\can}).\label{eq:canonical-identity}
\end{equation}
This identity does not require any of the locality or operational-time constraints, including the locality of feedback dependence on measurement outcomes.
Here, we have introduced the quantum relative entropy $D(\rho\Vert\sigma)=\Tr\!\left[\rho(\log\rho-\log\sigma)\right].
$
In particular, $\beta^{-1}D(\rho_f^\can\Vert\rho_i^{\can})$ serves as the nonequilibrium free energy of $\rho_f^{\can}$~\cite{Esposito2011}.
We have also defined
the quantum--classical (QC) mutual information~\cite{QC1,QC2,QC3,sagawa-ueda2008PRL} and the Holevo information~\cite{holevo1,holevo2,holevo3}
by using the von Neumann entropies $S(\rho)=-\mathrm{Tr}[\rho\log\rho]$ of the branch states $\rho_{\bm k}^{\prime\:\can}:=U_{\bm k}M_{\bm k}U_1\rho_i^{\rm ens}U_1^\dagger
M_{\bm k}^\dagger U_{\bm k}^\dagger/P_{\bm k}^{\rm ens}$:
\begin{align}
    I_{\QC}&:=S(\rho_i^{\can})-\sum_{\bm k}P_{\bm k}^{\can}S(\rho_{\bm k}^{\prime\:\can})\label{eq:IQC},\\
 \chi&:=S(\rho_f^{\can})-\sum_{\bm k}P_{\bm k}^{\can}S(\rho_{\bm k}^{\prime\:\can}).
 \label{eq:Holevo}
\end{align}
Because the feedback unitaries preserve the entropy of each conditional branch, $I_{\QC}$ is independent of $\{U_{\bm k}\}$ and quantifies the average entropy reduction obtained by conditioning on the measurement outcome, namely, the information acquired by the measurement. 
By contrast, 
$\chi$ quantifies how much information about the measurement outcome remains encoded in the final state after feedback.
In particular, $\chi=0$ if all postfeedback branches are identical, whereas $\chi$ becomes large when different outcomes lead to more distinguishable final states.
Also, $I_{\QC}-\chi$ represents the net entropy reduction.

Equation~\eqref{eq:canonical-identity} is therefore a thermodynamic
identity for a closed quantum many-body system, in which no heat bath is
involved. It shows that
the information $I_{\QC}$ acquired by the measurement is not converted
directly into work. The extractable work is necessarily reduced by the Holevo
information $\chi$ and the relative entropy
$D(\rho_f^\can\Vert\rho_i^{\can})$ (i.e., $\beta W_\can <I_{\QC}$ if $W_\can>0$). This differs from the setting of Ref.~\cite{sagawa-ueda2008PRL}, which considers contact with a heat bath. A detailed derivation of Eq.~\eqref{eq:canonical-identity}, as well as the information-thermodynamic meaning of 
the optimal extractable work $W_\can^\opt$ based on Eq.~\eqref{eq:canonical-identity}, is given in End Matter.

\PRLsection{Example}
We now explicitly calculate the extractable work and illustrate the physical meaning of Eq.~\eqref{eq:main-equivalence} for a simple setup. We consider a noninteracting $V$-qubit system
$ H=g\sum_{j=1}^{V}\sigma_j^3\:(g>0),
 $
where $\sigma_j^3$ is the Pauli operator acting only on site $j$:
$
\sigma_j^3=\sigma^0\otimes\sigma^0\otimes\cdots\otimes\sigma^0\otimes\sigma^3\otimes\sigma^0\otimes\cdots\otimes\sigma^0.
$
The corresponding canonical ensemble is given by
$
 \rho_i^{\can}=\left[
 \frac{\ee^{-\beta g\sigma^3}}{2\cosh(\beta g)}
 \right]^{\otimes V}.
$
We set the premeasurement unitary to $U_1=\id$. We perform the noisy local measurement on all qubits, whose strength is parametrized by $\eta~(0\le\eta\le1)$~\cite{Mochizuki-Hamazaki-MIST}:
\begin{align}
 M_{\bm k}&=\bigotimes_{j=1}^{V}m_{k_j}(\eta),\quad
 m_{k}(\eta)=
 \frac{\sigma^0+k\eta\sigma^3}{\sqrt{2(1+\eta^2)}},
 \quad k=\pm 1.
 \label{eq:qubit-measurement}
\end{align}
The variable $k_j$ is the outcome at site $j$ of the $\sigma^3$ measurement and takes the value $+$ or $-$. The case $\eta=1$ corresponds to a projective measurement, whereas $\eta=0$ corresponds to a measurement that yields no information.
Note that, in this case, the depth is $d_{\mathrm M}=1$, $Y$ is a single site, and $\mathcal{F}_{n(=1)}$ is the collection of all single-site regions.

We also restrict the feedback to product unitaries as
$
U_{\bm k}=\bigotimes_{j=1}^V u_{k_j},
$
where $u_{k_j}$ acts only on site $j$ and depends only on the
measurement outcome obtained at that site ($Z$ is a single site, and $\ell=0$). 
We then define the maximally extractable work under these single-qubit operations as
$W^{\rm loc-opt}
 =\max_{\{u_{k_j}\}}W
$
for MITE
and likewise $W^{\rm loc-opt}_\mathrm{ens}$ for the canonical ensemble.

First, we find (see the Supplemental Material for details \cite{supplement})
\begin{equation}
 W_{\can}^{\rm loc-opt}
 =Vg\max\!\left[
 0,\frac{2\eta}{1+\eta^2}-\tanh(\beta g)
 \right].
 \label{eq:local-optimal-work}
\end{equation}
 Equation~\eqref{eq:local-optimal-work} shows that the maximum extractable work has a nonanalytic onset at a critical inverse temperature 
$
\beta_c=g^{-1}\mathrm{arctanh}\left(\frac{2\eta}{1+\eta^2}\right) .
$
This nonanalytic behavior arises because the optimal feedback operation switches at that boundary, which is regarded as a control-landscape transition~\cite{Beato2025}. We can also verify that Eq.~\eqref{eq:canonical-identity} holds in this setting (see the Supplemental Material for details \cite{supplement}).

Next, consider work extraction from MITE.
Since the measurements act on individual sites and the feedback
at each site depends only on the measurement outcome obtained
there, the protocol satisfies the locality conditions above.
We can therefore apply Eq.~\eqref{eq:main-equivalence}:
\begin{equation}
W^{\loc-\opt}\simeq W_{\can}^{\rm loc-opt}.
 \label{eq:MITE-local-optimum}
\end{equation}
Thus, the work $W^{\loc-\opt}$ extractable from MITE also exhibits a nonanalytic onset at a parameter boundary, as in the canonical case,  in the thermodynamic limit.
Because the feedback Hamiltonian in this example contains only single-site terms, the corresponding Lieb-Robinson velocity is zero. Therefore,
no condition need be imposed on the feedback time $t$~\cite{supplement}.

Finally, since $  W_{\rm ens}^{\loc-\opt}\le W_{\can}^{\opt}$ by definition, where $W_{\can}^{\opt}$ allows arbitrary feedback unitaries, we obtain
\begin{equation}
  W^{\loc-\opt}\leq W_{\can}^{\opt}
  \label{eq:locanddae}
\end{equation}
from Eq.~\eqref{eq:MITE-local-optimum}. Here, we neglect the subextensive correction.
This is a manifestation of  \eqref{eq:daemonic-bound} in the current setup.
Interestingly, the equality condition for the inequality \eqref{eq:locanddae} is achieved with $\beta=0$ in our setting, as shown in End Matter.
In that sense, inequality~\eqref{eq:daemonic-bound}, which bounds the extractable work from MITE under these locality constraints by the optimal extractable work for the canonical ensemble, is tight.

\PRLsection{Anomalous work extraction from long-range correlations using distant measurement outcomes}
\label{sec:unrestricted-record}

We next relax the condition in
Eq.~\eqref{eq:feedback-outcome-locality} and demonstrate that feedback
based on distant measurement outcomes can exploit long-range
correlations hidden in MITE to extract an extensive amount of additional
work relative to the corresponding canonical ensemble.
We again consider the noninteracting Hamiltonian introduced above, assuming that the system is arranged in a one-dimensional chain $V=L$ and $L$ satisfies $L=3n$ with $n\in\mathbb{N}$.
As an initial state, we consider a product of GHZ triplets between distant sites:
\begin{align}
 |\Psi_0\rangle
 =\bigotimes_{j=1}^{n}
 \frac{|\uparrow\uparrow\uparrow\rangle_{j,j+n,j+2n}
       +|\downarrow\downarrow\downarrow\rangle_{j,j+n,j+2n}}{\sqrt2}.
 \label{eq:nonlocal-record-MITE}
\end{align}
This state satisfies the MITE condition with respect to the
canonical ensemble $\rho_i^{\can}=\id/2^L$ at $\beta=0$, while it has a long-range entanglement.
Indeed, any region $A$ with $\diam(A)<L/3$ contains at most
one site from each GHZ triplet, so its reduced density matrix
is exactly $\id_A/2^{|A|}$.

We set $U_1=\id$ and perform the noisy local measurement~\eqref{eq:qubit-measurement} on all sites.
For each $j$, we look at the outcomes $(k_j,k_{j+n},k_{j+2n})$, and if the number of $+1$ outcomes is larger than that of $-1$ outcomes among the three, we apply the unitary generated by
$
\sigma_j^1+\sigma_{j+n}^1+\sigma_{j+2n}^1.
$

At $t=\pi/2$, the straightforward calculation leads to
\begin{align}
 W=\frac{gL(3q-q^3)}{2},\quad
 W_{\can}=\frac{ gLq}{2},\quad
 W_{\can}^\mathrm{opt}=gLq,
 \label{eq:nonlocal-work-main}
\end{align}
where $q=\frac{2\eta}{1+\eta^2}$. Note that while $W_{\can}$ follows the same feedback protocol as the MITE, $W_{\can}^\mathrm{opt}$ allows arbitrary feedback protocols from the canonical ensemble.

We have $W> W_{\can}^\mathrm{opt}>W_{\can}$ for $0<\eta<1$.
Therefore, the measurement-and-feedback protocol extracts an extensive amount of additional work from $|\Psi_0\rangle$ compared with the corresponding canonical ensemble, exploiting the hidden long-range correlations associated with the entanglement structure of this pure MITE state.
This example shows that Eq.~\eqref{eq:main-equivalence} can fail even for \textit{constant-time} feedback generated by on-site terms if the feedback depends on distant measurement outcomes. In the End Matter, we summarize the relation between the work extraction and the operational time.

\PRLsection{Conclusion and outlook}
We have discussed information thermodynamics from possibly pure thermal
states by measurement and feedback operations in closed quantum
many-body systems,
when the measurements are local and the unitary controls are generated
by local Hamiltonians over finite times (for $s=0$).
We have first shown that, when each local
feedback term depends only on nearby measurement outcomes, the work extractable from MITE is approximated by that obtained by applying the same protocol to a corresponding canonical ensemble.
We have then expressed the extractable work in terms of the information-theoretical quantities. 

For a noninteracting $V$-qubit system, we have demonstrated that the maximum work $W^{\loc-\opt}$ extractable from MITE under these locality constraints is bounded by the optimal extractable work
$W_{\can}^{\opt}$ for a canonical initial state, and that the equality condition is achieved at the infinite temperature.
We have also shown that long-range correlations in a pure MITE
state can be exploited through feedback based on distant
measurement outcomes, yielding an extensive amount of additional
work relative to the corresponding canonical ensemble.

There are several future directions.
First, it would be interesting to investigate how to relate the additional work gain from the quantum correlation to the known entanglement measures.
It would also be interesting to investigate how recent LOCC-based approaches to characterizing many-body quantum correlations through work extraction~\cite{YYS2025,YYS2026} relate to our framework in interacting closed quantum many-body systems.
Another intriguing direction is to extend our results to repeated and continuous measurement and feedback~\cite{Horowitz2010,Munakata2014,Yada2022}.
Finally, it is of importance to investigate the nature of control-landscape transitions induced by measurement and feedback, found in noninteracting systems in this work, in more complicated interacting systems.

\PRLsection{Acknowledgments}
We would like to thank Ken Funo, Yuuya Chiba, and Kohei Yoshimura for fruitful discussions.
OpenAI's ChatGPT (GPT-6) was used to find the example of the work extraction from GHZ states and to assist with the development and checking of parts of the derivation of the MITE--canonical equivalence. The AI-assisted mathematical arguments included in the manuscript were critically reviewed and independently verified by the authors.
We are supported by JST ERATO Grant No.~JPMJER2302, Japan.
S.S. was supported by KAKENHI Grant No.\ JP26K17047.
R.H. was supported by KAKENHI Grant No.\ JP24K16982.

\bibliography{bib}

\clearpage

\section{End Matter}

\PRLsection{Derivation of the information-thermodynamic identity from the canonical state}
We give the details of the derivation of Eq.~\eqref{eq:canonical-identity}.
First, the relative entropy between the final state $\rho_f^\can$  and the initial state $\rho_i^{\rm ens}$ is rewritten as
\begin{align}
    D(\rho_f^\can||\rho_{i}^{\rm ens})&={\rm Tr}[\rho_f^\can \log \rho_f^\can]-{\rm Tr}[\rho_f^\can \log \rho_{i}^{\rm ens}]\nonumber\\
    &=-S(\rho_f^\can)-{\rm Tr}[\rho_f^\can \left(-\beta H-\log Z\right)]\nonumber\\
    &=-S(\rho_f^\can)+\beta{\rm Tr}[H \rho_f^\can]+\log Z.\label{relative_en}
\end{align}
Thus,
\begin{align}
    \beta{\rm Tr}[H \rho_f^\can]=D(\rho_f^\can||\rho_{i}^{\rm ens})+S(\rho_f^\can)-\log Z.\label{fene}
\end{align}
Moreover,
\begin{align}
    \beta{\rm Tr}[H \rho_i^{\rm ens}]=-\log Z+S(\rho_i^{\rm ens}).\label{iene}
\end{align}
From Eqs.~\eqref{fene} and \eqref{iene}, the extractable work \eqref{eq:work-definition} from the canonical state is
\begin{align}
    \beta W_{\can}
    &=S(\rho_i^{\rm ens})-S(\rho_f^\can)-D(\rho_f^\can||\rho_{i}^{\rm ens})\nonumber\\
    &=S(\rho_i^{\can})-
 \sum_{\bm k}P_{\bm k}^{\can}S(\rho_{\bm k}^{\prime\:\can})\notag\\
    &-\left[S(\rho_f^\can)-
 \sum_{\bm k}P_{\bm k}^{\can}S(\rho_{\bm k}^{\prime\:\can})\right]-D(\rho_f^\can||\rho_{i}^{\rm ens})\nonumber\\
 &=I_{\QC}-\chi-D(\rho_f^\can\Vert\rho_i^{\can}).
\end{align}
Thus, it can be expressed in terms of the quantum--classical (QC) mutual information \eqref{eq:IQC}, the Holevo information \eqref{eq:Holevo}, and the relative entropy.
Note that this equality becomes trivial (right- and left-hand sides become zero) for $\beta =0$, although Eq.~\eqref{eq:main-equivalence} nontrivially holds even for this case.

\PRLsection{Information theoretical meaning of $W_\can^\opt$}
Equation~\eqref{eq:canonical-identity} also clarifies the information-thermodynamic meaning of 
the optimal extractable work $W_\can^\opt$.
We denote the conditional state immediately after the measurement and before feedback by
$
 \widetilde{\rho}_{\bm k}^{\can}=M_{\bm k}U_1\rho_i^{\can}U_1^\dagger M_{\bm k}^\dagger/P_{\bm k}^{\can}.
$
Let $H\ket{E_{\alpha}}=E_{\alpha}\ket{E_{\alpha}}$ with $E_1\le E_2\le\cdots$, and let $r_{\bm k,1}\ge r_{\bm k,2}\ge\cdots$ be the eigenvalues of $\widetilde{\rho}_{\bm k}^{\can}$.
Since a feedback unitary preserves these eigenvalues, the branch energy is minimized by assigning the largest eigenvalues to the lowest energy levels.
Thus, if arbitrary feedback unitaries are allowed, the optimal postfeedback branch is
$
 \rho_{\bm k}^{\prime\:\can,\opt}=\sum_{\alpha} r_{\bm k,\alpha}\ket{E_{\alpha}}\bra{E_{\alpha}}.
$
This state is passive.
Applying Eq.~\eqref{eq:canonical-identity} to this optimal feedback gives
\begin{equation}
 \beta W_{\can}\leq\beta W_{\can}^{\opt}=I_{\QC}-\chi^{\opt}- D(\rho_f^{\can,\opt}\Vert\rho_i^{\can})
\end{equation}
with $\rho_f^{\can,\opt}=\sum_{\bm{k}}P_{\bm k}^{\can}\rho_{\bm k}^{\prime\:\can,\opt}$.
Here, $I_{\QC}$ is unchanged by the feedback optimization, whereas $\chi^{\opt}$ and $\rho_f^{\can,\opt}$ are evaluated for the optimal feedback.
Hence, Eq.~\eqref{eq:canonical-identity} provides an information-thermodynamic decomposition of the optimal extractable work in a closed quantum system.
Interestingly, $I_{\QC}$, $\chi^{\opt}$ and $D(\rho_f^{\can,\opt}\Vert\rho_i^{\can})$ are expressed by classical populations (e.g., $r_{\bm{k},n}$) alone, since $\rho_f^{\can,\opt}$ and $\rho_i^{\can}$ are both diagonalized in the energy basis.

\PRLsection{Equality condition in Eq.~\eqref{eq:locanddae}}
Here, we show that the equality condition in Eq.~\eqref{eq:locanddae} is achieved for $\beta=0$ in the setup of noninteracting $V$ qubits.
Since $W^{\loc-\opt}\simeq W^{\loc-\opt}_{\can}$, it is enough to show $W^{\loc-\opt}_{\can}=W^{\opt}_{\can}$.
For this purpose, we will show that 
 the product feedback operation described by $U_{\bm k}=\bigotimes_{j=1}^V u_{k_j}$ already makes every conditional branch passive at $\beta=0$.
Then, allowing feedback operations that create correlations between qubits cannot lower the energy any further.

The initial canonical state is the infinite-temperature state,
$\rho_i^{\can}=(\sigma^0/2)^{\otimes V}$.
After obtaining the single-qubit outcome $k=\pm1$, the normalized postmeasurement state for each qubit reads
\begin{equation}
 \frac{m_k(\eta)(\sigma^0/2)m_k^\dagger(\eta)}{\Tr[m_k(\eta)(\sigma^0/2)m_k^\dagger(\eta)] }=
 \frac{\sigma^0+k q\sigma^3}{2},
 \qquad
 q=\frac{2\eta}{1+\eta^2}.
 \label{eq:beta0-single-qubit-branch}
\end{equation}
Equation~\eqref{eq:beta0-single-qubit-branch} shows that, conditioned on the outcome, the measurement prepares a qubit biased along the $\sigma^3$ axis: the $k=+1$ branch is biased toward $\sigma^3=+1$, whereas the $k=-1$ branch is biased toward $\sigma^3=-1$. The locally optimal feedback at $\beta=0$, $u_-=\sigma^0$ and $u_+=\sigma^1$, then uses the outcome only to orient this bias toward the lower-energy direction:
\begin{equation}
 u_k \left(\frac{\sigma^0+k q\sigma^3}{2}\right)u_k^\dagger
 =
 \frac{\sigma^0-q\sigma^3}{2}.
 \label{eq:beta0-local-rectification}
\end{equation}
Consequently, for every outcome $\bm k$, the normalized state after measurement and feedback is
\begin{align}
 \rho_{\bm k}^{\prime\:\can}&=\left(\frac{\sigma^0-q\sigma^3}{2}\right)^{\otimes V}\notag\\
 &=\frac{\exp[-a\sum_{j=1}^{V}\sigma_j^3]}{(2\cosh a)^V}=\frac{\exp[-(a/g)H]}{(2\cosh a)^V},
 \label{eq:beta0-passive-branch}
\end{align}
where $a=\operatorname{arctanh}q$. Equation~\eqref{eq:beta0-passive-branch} shows that each conditional branch is already a canonical state of the full Hamiltonian at a nonnegative inverse temperature $a/g$.
Therefore it is passive. 
Any further unitary feedback, including an entangling one, cannot lower the energy once the branch has already been put in this passive state.
Hence the equality in Eq.~\eqref{eq:locanddae} holds when $\beta=0$.

\begin{table*}
\caption[Operation]{\label{tbl:Comparison}Summary of the operational time and extractable work  from the MITE in closed quantum many-body systems. }
\begin{ruledtabular}
\begin{tabular}{lll}
    Processes & Constant times\quad & Long times\\
    \colrule
  No measurement and feedback~\cite{Hokkyo2025}   & $W\lesssim 0$ & $W>0$ is possible~\cite{Chiba2026}  \\
Feedback based on nearby measurement outcomes [this work]  & $W\lesssim W_\can^\mathrm{opt}$ [Eq.~\eqref{eq:daemonic-bound}] & $W> W_\can^\mathrm{opt}$ is possible\\
Feedback based on distant measurement outcomes [this work]  & $W> W_\can^\mathrm{opt}$ is possible [Eq.~\eqref{eq:nonlocal-work-main}] & $W> W_\can^\mathrm{opt}$ is possible \\
\end{tabular}
\end{ruledtabular}
\end{table*}

\PRLsection{Summary of the operational time and extractable work}
In Table~\ref{tbl:Comparison}, we summarize the relation between the operational time and extractable work  from MITE with $s=0$ for different protocols, i.e., (a) protocol with no measurement and feedback, (b) feedback-based protocol using nearby measurement outcomes, and (c) feedback-based protocol using distant measurement outcomes.
We assume that the unitary operations are generated by local Hamiltonians, and, if any,  measurement is composed of local gates.
We consider two regimes, i.e., constant times (e.g., $t=\mathcal{O}(L^0)$) and longer times  (e.g., $t=\mathcal{O}(L)$).
In the case (b), we can obtain positive extensive work prohibited in the case (a) even for constant times, although it is still bounded by the optimal work predicted by the canonical ensemble.
In case (c), we can use long-range correlation to obtain additional work extensively larger than that from the canonical ensemble, even for the constant times.


\widetext

\clearpage

\begin{center}
{\bf \large Supplemental Material}
\end{center}

\renewcommand{\theequation}{S\arabic{equation}}

\renewcommand{\thefigure}{S\arabic{figure}}
\renewcommand{\thetable}{S\arabic{table}}

\setcounter{equation}{0}
\setcounter{figure}{0}
\setcounter{table}{0}

\section{S1. Derivation of the MITE--canonical equivalence}

\begin{figure}[tp]
 \begin{center}
  \includegraphics[scale=0.09]{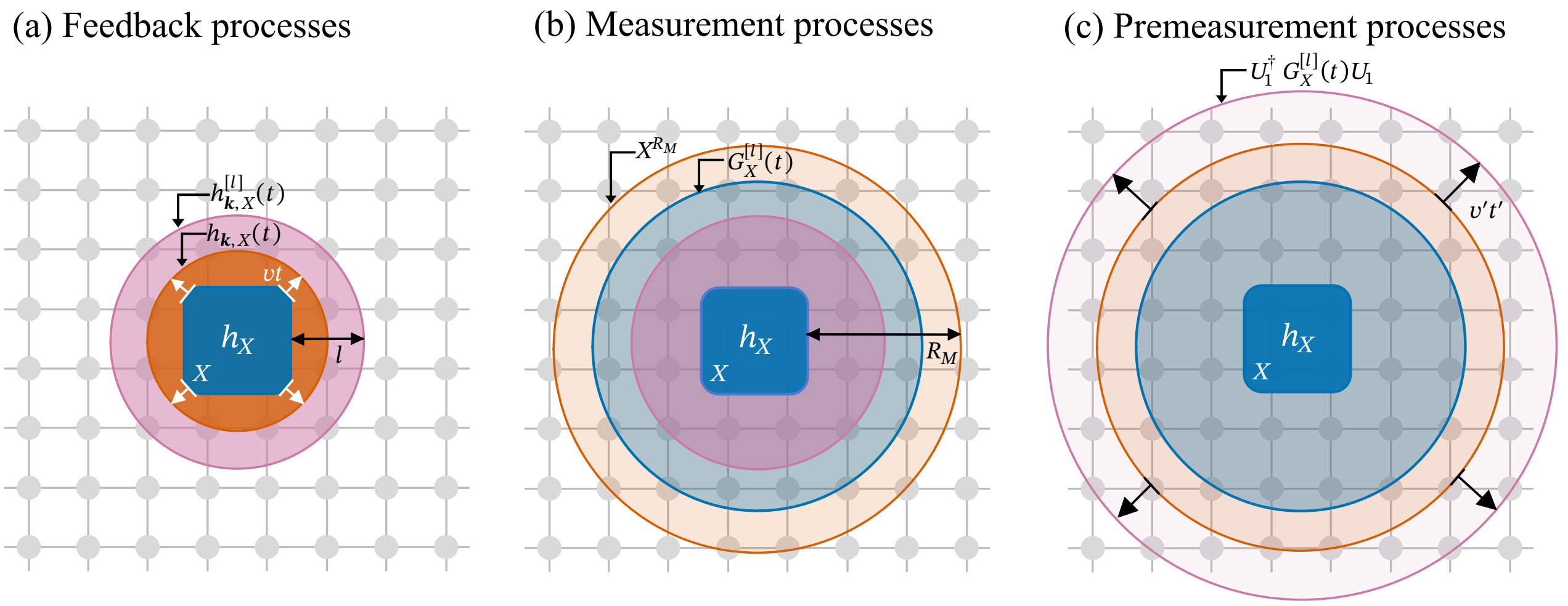}
\caption{Schematic illustration of operator spreading.
(a) A local operator $h_X$, initially supported on $X$,
spreads under the feedback unitary $U_{\bm k}(t)$,
but the Lieb--Robinson bound allows it to be approximated
by $h_{\bm k,X}^{[l]}(t)$, defined using the truncated
Hamiltonian $H_{\bm k}^{X^l}$ and supported on $X^l$,
with an exponentially small error when $l-r_0-vt$
is sufficiently large.
(b) Under the locality condition in Eq.~\eqref{feedcon},
summing over the outcomes of the depth-$d_{\mathrm M}$
local measurement circuit $M_{\bm k}$ gives
$G_X^{[l]}(t):=\sum_{\bm k}M_{\bm k}^{\dagger}
h_{\bm k,X}^{[l]}(t)M_{\bm k}$, supported on a region
${\rm supp}(G_X^{[l]}(t))\subseteq X^{R_M}$, where
$R_M=l+\ell+(\mathscr D-1)d_{\mathrm M}$.
(c) Finally, the time evolution under the premeasurement
unitary $U_1(t')$ gives rise to an additional
Lieb--Robinson light cone.
Consequently, the outcome-averaged operator
$\sum_{\bm k}U_1^\dagger(t')M_{\bm k}^\dagger
U_{\bm k}^\dagger(t)h_XU_{\bm k}(t)
M_{\bm k}U_1(t')$
can be approximated by an operator supported on $X^{l'}$,
with an exponentially small error when both $l-r_0-vt$
and $l'-R_M-v't'$ are sufficiently large.
\label{speading}}
  \end{center}
\end{figure}

We give the details of the derivation of Eq.~\eqref{eq:main-equivalence}.
Assume that the initial state is a MITE state $\rho_{i}^{\MITE}$ and that the Hamiltonian $H=\sum_Xh_X$ is a sum of local observables.
Then, for the extractable work
\begin{align}
    W={\rm Tr}[H\rho_{i}^{\MITE}]-{\rm Tr}[H\rho_{f}],\label{10-1}
\end{align}
the first term satisfies
\begin{align}
    {\rm Tr}[H\rho_i^{\MITE}]\simeq{\rm Tr}[H\rho_i^{\rm ens}]\label{eq:initial-energy-equivalence}
\end{align}
by the definition of MITE.
The nontrivial contribution is the second term, namely, the final-state energy. For the initial state $\rho_i^{\MITE}$, the second term in Eq.~\eqref{10-1} can be written as
\begin{align}
 \Tr[H\rho_f]
 =\sum_{X,\bm k}\Tr\!\left[
 \rho_i^{\MITE} U_1^\dagger M_{\bm k}^\dagger
 U_{\bm k}^\dagger(t)h_XU_{\bm k}(t)
 M_{\bm k}U_1\right].
 \label{eq:final-energy-Heisenberg}
\end{align}
Because $\rho^{\MITE}_i$ is MITE, if $\sum_{\bm k}U_1^{\dagger}M_{\bm k}^{\dagger}U_{\bm k}(t)^{\dagger}h_{X}U_{\bm k}(t)M_{\bm k}U_1$ is approximated by an $\mathcal{O}(L^s)$-local observable, then
\begin{align}
    &\sum_{\bm{k}}\sum_{X}{\rm Tr}[\rho_{i}^{\MITE}U_1^{\dagger}M_{\bm k}^{\dagger}U_{\bm k}(t)^{\dagger}h_{X}U_{\bm k}(t)M_{\bm k}U_1]\notag\\
    &\simeq\sum_{\bm{k}}\sum_{X}{\rm Tr}[\rho_{i}^{\rm ens}U_1^{\dagger}M_{\bm k}^{\dagger}U_{\bm k}(t)^{\dagger}h_{X}U_{\bm k}(t)M_{\bm k}U_1].
\end{align}
This establishes Eq.~\eqref{eq:main-equivalence}. It is therefore sufficient to show that
\begin{align}
    \sum_{\bm k}U_1^{\dagger}M_{\bm k}^{\dagger}U_{\bm k}(t)^{\dagger}h_{X}U_{\bm k}(t)M_{\bm k}U_1\label{shouldbeproved}
\end{align}
is well approximated by an $\mathcal{O}(L^s)$-local observable.
Although this operator appears nonlocal, the Lieb-Robinson bound controls the error in a local approximation, as explained below~(see also Fig. \ref{speading}).
In the following, we assume that $H_1(t)$ and $H_{\bm k}(t)$ are time-independent for simplicity.
Note that the extension to time-dependent cases is straightforward; as pointed out in note [12] of Ref.~\cite{Bravyi2006}, the Lieb--Robinson bound also applies to time-dependent local Hamiltonians, provided that the local interaction strengths are uniformly bounded.

\subsection{Feedback processes~(Fig. \ref{speading}(a))}
In the following, we assume that the operator norm of $h_X$ is bounded by a finite value independent of the system size $V$.
First, we set
\begin{align}
    U_{\bm k}^{\dagger}(t)h_X U_{\bm k}(t):=h_{{\bm k},X}(t),
\end{align}
and evaluate
\begin{align}
    \left|\sum_{\bm{k}}\sum_{X}{\rm Tr}[\rho_{i}^{\MITE}U_1^{\dagger}M_{\bm k}^{\dagger}h_{{\bm k},X}(t)M_{\bm k}U_1]-\sum_{\bm{k}}\sum_{X}{\rm Tr}[\rho_{i}^{\MITE}U_1^{\dagger}M_{\bm k}^{\dagger}h_{{\bm k},X}^{[l]}(t)M_{\bm k}U_1]\right|.\label{13-1}
\end{align}
To explain the definition of $h_{{\bm k},X}^{[l]}(t)$,
we now write the feedback Hamiltonian as
\begin{align*}
    H_{\bm k}=\sum_Z\tilde h_{\bm k^{(Z)},Z}.
\end{align*}
Each term $\tilde h_{\bm k^{(Z)},Z}$ is a local operator supported on a region $Z$ with $\operatorname{diam}(Z)\le r_0$, which is independent of $L$.
Also, $\tilde h_{\bm k^{(Z)},Z}$ depends on $\bm k^{(Z)}$, which denotes a set of the measurement outcomes $k_{n,Y}$ associated with the region $Z$.
We impose the following locality condition on  $\bm k^{(Z)}$, which indicates that the feedback Hamiltonian depends on nearby measurement outcomes.
Specifically, to select $k_{n,Y}$ in $\bm k^{(Z)}$, we assume that the corresponding measurement region $Y$ must contain at least one site $y_0$ such that
\begin{align}
    {\rm dist}(Z,y_0)\le\ell,
    \qquad \ell=\mathcal O(L^s)\label{feedcon}
\end{align}
for a fixed $s\:(0\leq s<1)$.

We now define $h_{{\bm k},X}^{[l]}(t)$.
Let $X^l:=\{x:\operatorname{dist}(x,X)<l\}$ with $l>0$.
We retain only the Hamiltonian terms whose regions $Z$ are entirely contained in $X^l$:
\begin{align*}
    H_{\bm k}^{X^l}&:=\sum_{Z\subseteq X^l}\tilde h_{\bm k^{(Z)},Z},\\
    U_{\bm k}^{X^l}(t)&:=\Torder\exp\!\left[-\ii\int_0^t H_{\bm k}^{X^l}\,du\right].
\end{align*}
We use this restricted evolution to define the local approximation
\begin{align}
    h_{{\bm k},X}^{[l]}(t):=
    \bigl(U_{\bm k}^{X^l}(t)\bigr)^\dagger h_X U_{\bm k}^{X^l}(t).
    \label{30-1}
\end{align}
Since every retained term is supported inside $X^l$, on the full Hilbert space we can write
\begin{align*}
    H_{\bm k}^{X^l}
    &=K_{\bm k,X^l}\otimes\id_{\overline{X^l}},\\
    U_{\bm k}^{X^l}(t)
    &=e^{-\ii tK_{\bm k,X^l}}\otimes\id_{\overline{X^l}},
\end{align*}
where $K_{\bm k,X^l}$ acts on the Hilbert space of $X^l$.

We now estimate the error in Eq.~(\ref{13-1}).
We define
\begin{align}
    B_{{\bm k},X}:=h_{{\bm k},X}(t)-h_{{\bm k},X}^{[l]}(t).
\end{align}
Then, Eq.~(\ref{13-1}) can be rewritten as follows: 
\begin{align}
    {\rm Eq.~(\ref{13-1})}&=\left|\sum_{X}{\rm Tr}[\sum_{\bm{k}}M_{\bm k}U_1\rho_{i}^{\MITE}U_1^{\dagger}M_{\bm k}^{\dagger}B_{{\bm k},X}]\right|\nonumber\\
    &=\left|\sum_{X}{\rm Tr}[\sum_{\bm{k}}P_{\bm k}\frac{M_{\bm k}U_1\rho_{i}^{\MITE}U_1^{\dagger}M_{\bm k}^{\dagger}}{P_{\bm k}}B_{{\bm k},X}]\right|,\label{14-1}
\end{align}
where
\begin{align}
P_{\bm k}
 &=\Tr\!\left[M_{\bm k}U_1\rho_i^{\MITE}U_1^\dagger M_{\bm k}^\dagger\right].
\end{align}
We further define
\begin{align}
    \widetilde{\rho}_{\bm k}:=\frac{M_{\bm k}U_1\rho_{i}^{\MITE}U_1^{\dagger}M_{\bm k}^{\dagger}}{P_{\bm k}}.
\end{align}
Then,
\begin{align}
    {\rm Eq.~(\ref{14-1})}&=\left|\sum_{X}{\rm Tr}[\sum_{\bm{k}}P_{\bm k}\widetilde{\rho}_{\bm k}B_{{\bm k},X}]\right|\notag\\
    &=\left|\sum_{X}\sum_{\bm{k}}P_{\bm k}{\rm Tr}[\widetilde{\rho}_{\bm k}B_{{\bm k},X}]\right|\notag\\
    &\le\left|\sum_{X}\max_{\bm k}|{\rm Tr}[\widetilde{\rho}_{\bm k}B_{{\bm k},X}]|\:\right|\notag\\
     &\le\sum_{X}\max_{\bm k}\|B_{{\bm k},X}\|_{\infty}\notag\\
    &=\sum_{X}\max_{\bm k}\|h_{{\bm k},X}(t)-h_{{\bm k},X}^{[l]}(t)\|_{\infty},\label{15-1}
\end{align}
where $\|\cdots\|_\infty$ denotes the operator norm.
Since $h_{\bm k,X}^{[l]}(u)$ is supported on $X^l$,
it commutes with every Hamiltonian term lying entirely
outside $X^l$.

For the time-independent Hamiltonians $H_{\bm k}^{X^l}$ considered here,
the Heisenberg equation gives
\begin{align*}
&\frac{d}{du}h_{\bm k,X}^{[l]}(u)\\
&=\frac{d}{du}\left[\bigl(U_{\bm k}^{X^l}(u)\bigr)^\dagger h_X U_{\bm k}^{X^l}(u)\right]\\
&=\ii[H_{\bm k}^{X^l},\bigl(U_{\bm k}^{X^l}(u)\bigr)^\dagger h_X U_{\bm k}^{X^l}(u)]\\
&\left({\rm \because\ \  }U_{\bm k}^{X^l}(t):=\Torder\exp\!\left[-\ii\int_0^t H_{\bm k}^{X^l}\,du\right]\right)\\
&=\ii[H_{\bm k}^{X^l},h_{\bm k,X}^{[l]}(u)]. 
\end{align*}
Differentiating the following product therefore yields
\begin{align*}
    &\frac{d}{du}\left[
    U_{\bm k}^\dagger(t-u)
    h_{\bm k,X}^{[l]}(u)
    U_{\bm k}(t-u)\right]\\
    &\quad=
    -\ii\,U_{\bm k}^\dagger(t-u)
    \left[H_{\bm k}-H_{\bm k}^{X^l},
    h_{\bm k,X}^{[l]}(u)\right]
    U_{\bm k}(t-u).
\end{align*}
Integrating from $0$ to $t$, taking the operator norm,
and using its invariance under unitary operations,
we obtain
\begin{align}
    &\bigl\|h_{\bm k,X}(t)
    -h_{\bm k,X}^{[l]}(t)\bigr\|_\infty\notag\\
    &\quad\le
    \int_0^t
    \bigl\|[H_{\bm k}-H_{\bm k}^{X^l},
    h_{\bm k,X}^{[l]}(u)]\bigr\|_\infty\,du\label{tochu1}
\end{align}
since
\begin{align*}
    &U_{\bm k}^\dagger(t-u)
    h_{\bm k,X}^{[l]}(u)
    U_{\bm k}(t-u)|_{u=t}=h_{\bm k,X}^{[l]}(t),\\
    &    U_{\bm k}^\dagger(t-u)
    h_{\bm k,X}^{[l]}(u)
    U_{\bm k}(t-u)|_{u=0}=h_{\bm k,X}(t).
\end{align*}
We now consider the integral $\int_0^t
    \bigl\|[H_{\bm k}-H_{\bm k}^{X^l},
    h_{\bm k,X}^{[l]}(u)]\bigr\|_\infty\,du$ in Eq.~(\ref{tochu1}).
The only terms that can fail to commute with $h_{\bm k,X}^{[l]}(u)$ are those supported entirely within $X^l$ or crossing the boundary of $X^l$.
By the definition of the truncated Hamiltonian,
$H_{\bm k}-H_{\bm k}^{X^l}
=\sum_{Z\not\subseteq X^l}\tilde h_{\bm k^{(Z)},Z}$,
so $H_{\bm k}-H_{\bm k}^{X^l}$ contains no terms supported entirely within $X^l$.
Thus, the only nonzero contributions to $[H_{\bm k}-H_{\bm k}^{X^l},
h_{\bm k,X}^{[l]}(u)]$ come from terms crossing the boundary of $X^l$.
Applying the triangle inequality to their sum gives
\begin{align*}
    &\bigl\|h_{\bm k,X}(t)
    -h_{\bm k,X}^{[l]}(t)\bigr\|_\infty\\
    &\quad\le
    \sum_{\substack{Z\cap X^l\neq\varnothing\\
                    Z\not\subseteq X^l}}
    \int_0^t
    \bigl\|[\tilde h_{\bm k^{(Z)},Z},
    h_{\bm k,X}^{[l]}(u)]\bigr\|_\infty\,du.
\end{align*}
For the evolution generated by the truncated Hamiltonian $H_{\bm k}^{X^l}$,
the Lieb--Robinson bound reads
\begin{align*}
    &\bigl\|[\tilde h_{\bm k^{(Z)},Z},
    h_{\bm k,X}^{[l]}(u)]\bigr\|_\infty\\
    &\quad\le c_{\rm LR}
    \|\tilde h_{\bm k^{(Z)},Z}\|_\infty\|h_X\|_\infty
    e^{-\kappa[\operatorname{dist}(X,Z)-vu]}.
\end{align*}
Here, $v\ge 0$ is a Lieb--Robinson velocity associated with the feedback Hamiltonian $H_{\bm k}^{X^l}$, and $\kappa>0$ is the spatial decay rate of the Lieb--Robinson tail. 
$c_{\rm LR}$, $v$ and $\kappa$ are independent of the system size $L$.
Note that this is the commutator form of Theorem~1 of
Ref.~\cite{BarthelKliesch2012Quasilocality} for Hamiltonian
dynamics; see also
Ref.~\cite{KlieschGogolinEisert2014LocalObservables}.
The factors depending on the uniformly bounded sizes
of $X$ and $Z$ are included in $c_{\rm LR}$.

For each such region $Z$, choose a site
$z_{\rm out}\in Z\setminus X^l$.
By the definition of $X^l$,
$\operatorname{dist}(z_{\rm out},X)\ge l$.
The condition $\operatorname{diam}(Z)\le r_0$ means
that the distance between any two sites in $Z$ is
at most $r_0$.
Since $z_{\rm out}$ is a site in $Z$, we have
$\operatorname{dist}(z,z_{\rm out})\le r_0$
for every site $z\in Z$.
The triangle inequality therefore gives
\begin{align*}
    \operatorname{dist}(z,X)
    &\ge \operatorname{dist}(z_{\rm out},X)
    -\operatorname{dist}(z,z_{\rm out})\\
    &\ge l-r_0.
\end{align*}
Taking the minimum over $z\in Z$, we obtain
$\operatorname{dist}(X,Z)\ge l-r_0>0$ for $l>r_0$.
Thus,
\begin{align*}
    &\bigl\|[\tilde h_{\bm k^{(Z)},Z},
    h_{\bm k,X}^{[l]}(u)]\bigr\|_\infty\\
    &\quad\le c_{\rm LR}
    \|\tilde h_{\bm k^{(Z)},Z}\|_\infty\|h_X\|_\infty
    e^{-\kappa(l-r_0)}e^{\kappa vu}.
\end{align*}

We next bound the sum of the norms of the Hamiltonian
terms crossing the boundary of $X^l$.
For a lattice site $x$, the sum $\sum_{Z\ni x}$ runs
over all local Hamiltonian terms whose regions $Z$
contain $x$.
On the fixed lattice, the bound
$\operatorname{diam}(Z)\le r_0$, with $r_0$ independent
of $L$, bounds the number of such terms independently
of $L$.
So, we assume that their norms are uniformly bounded, and 
$\sum_{Z\ni x}\|\tilde h_{\bm k^{(Z)},Z}\|_\infty$
has an upper bound independent of the system size.
We can therefore choose a finite constant $J$,
independent of $L$, $\bm k$, such that
\begin{align*}
    \sum_{Z\ni x}
    \|\tilde h_{\bm k^{(Z)},Z}\|_\infty\le J.
\end{align*}
We now sum this bound over the sites $x\in X^l$.
In this double sum, a term with region $Z$ is counted
once for each site in $Z\cap X^l$.
Every boundary-crossing region contains at least one
site in $X^l$, so each boundary term is counted at least
once.
Repeated counting and the inclusion of terms entirely
inside $X^l$ can only increase the sum, since all norms
are nonnegative.
Hence,
\begin{align*}
    &\sum_{\substack{Z\cap X^l\neq\varnothing\\
                    Z\not\subseteq X^l}}
    \|\tilde h_{\bm k^{(Z)},Z}\|_\infty\\
    &\quad\le
    \sum_{x\in X^l}\sum_{Z\ni x}
    \|\tilde h_{\bm k^{(Z)},Z}\|_\infty\\
    &\quad\le
    \sum_{x\in X^l}J
    =J|X^l|,
\end{align*}
where $|X^l|$ is the number of lattice sites in $X^l$.
Substituting these estimates into the preceding integral bound on the truncation error, we obtain
\begin{align*}
    &\bigl\|h_{\bm k,X}(t)
    -h_{\bm k,X}^{[l]}(t)\bigr\|_\infty\\
    &\quad\le c_{\rm LR}J\|h_X\|_\infty |X^l|
    e^{-\kappa(l-r_0)}
    \int_0^t e^{\kappa vu}\,du.
\end{align*}

We first estimate the number of sites in $X^l$.
On a $d$-dimensional hypercubic lattice,
all sites at distance less than $l$ from a given site $x$
lie inside a box centered at $x$ with side length $2l$
in each spatial direction.
The box therefore contains at most $(1+2l)^d$ lattice
sites.
Since one such box centered at each site of $X$
covers $X^l$, we obtain
\begin{align*}
    |X^l|
    \le |X|(1+2l)^d
    \le c_0(1+2l)^d.
\end{align*}
The constant $c_0$ is a uniform upper bound on $|X|$
and is independent of $L$.
Next, for $\kappa,v>0$ and $t\ge0$, the time integral is
\begin{align*}
    \int_0^t e^{\kappa vu}\,du
    &=\frac{e^{\kappa vt}-1}{\kappa v}\\
    &\le\frac{e^{\kappa vt}}{\kappa v}.
\end{align*}
Substituting both estimates into the preceding bound
gives, for each measurement record $\bm k$,
\begin{align*}
    &\bigl\|h_{\bm k,X}(t)
    -h_{\bm k,X}^{[l]}(t)\bigr\|_\infty\\
    &\quad\le
    \frac{c_{\rm LR}Jc_0}{\kappa v}
    \|h_X\|_\infty(1+2l)^d
    e^{-\kappa(l-r_0)}e^{\kappa vt}.
\end{align*}
Combining the exponential factors and setting
$C:=c_{\rm LR}Jc_0/(\kappa v)$, we obtain
\begin{align}
    &\max_{\bm k}
    \bigl\|h_{\bm k,X}(t)
    -h_{\bm k,X}^{[l]}(t)\bigr\|_\infty
    \notag\\
    &\qquad\le
    C\|h_X\|_\infty(1+2l)^d
    e^{-\kappa(l-r_0-vt)}.
    \label{15-2}
\end{align}

Combining Eqs.~\eqref{15-1} and \eqref{15-2}, we find that the total
feedback approximation error in Eq.~\eqref{13-1} obeys
\begin{align}
    \mathrm{Eq.}~\eqref{13-1}
    &\le C(1+2l)^d e^{-\kappa(l-r_0-vt)}
    \sum_X\|h_X\|_\infty.
    \label{15-20}
\end{align}
The sum $\sum_X\|h_X\|_\infty$ on the
right-hand side contains $\mathcal O(L^d)$ terms.
By assumption, each norm is bounded by the same constant,
independent of $L$.
The sum is therefore $\mathcal O(L^d)$.
Since $C$ is independent of $L$, $l$, and $t$, we obtain
\begin{align*}
    &C(1+2l)^d e^{-\kappa(l-r_0-vt)}
    \sum_X\|h_X\|_\infty\\
    &\qquad=
    \mathcal O(L^d)\,(1+2l)^d
    e^{-\kappa(l-r_0-vt)}.
\end{align*}
Therefore,
\begin{align}
    \mathrm{Eq.}~\eqref{13-1}\le
    \mathcal O(L^d)\,(1+2l)^d e^{-\kappa(l-r_0-vt)}.
    \label{bb}
\end{align}
It follows that, if $l$ satisfies
\begin{align}
    l=\mathcal{O}(L^s)~~~(0<s)~~\&~~t\ll\frac{l-r_0}{v},
\end{align}
where $t$ is the duration of the feedback control, then in the thermodynamic limit, 
the bound in Eq.~\eqref{bb} is exponentially suppressed.
Therefore, the operator $h_{{\bm k},X}(t)$ in the first term of Eq.~(\ref{13-1}) can be approximated by $h_{{\bm k},X}^{[l]}(t)$:
\begin{align}
    &\sum_{\bm k,X}
    \Tr\!\left[
    \rho_i^{\MITE}U_1^\dagger M_{\bm k}^\dagger
    h_{\bm k,X}(t)M_{\bm k}U_1
    \right]
    \notag\\
    &\qquad\simeq
    \sum_{\bm k,X}
    \Tr\!\left[
    \rho_i^{\MITE}U_1^\dagger M_{\bm k}^\dagger
    h_{\bm k,X}^{[l]}(t)M_{\bm k}U_1
    \right].
    \label{feed-eq}
\end{align}

The preceding argument assumes $s>0$ for $l=\mathcal{O}(L^s)$
and therefore does not directly apply to the case $s=0$, which corresponds to the ordinary MITE.
Nevertheless, Eq.~\eqref{feed-eq} remains valid for $s=0$, as we now show. 
For the work considered here, the notation $W\simeq W_{\can}$ means that the difference between $W$ and $W_{\can}$ is subextensive:
\begin{align}
    W\simeq W_{\can}\quad\Longleftrightarrow\quad|W-W_{\can}|=o(V)\quad\Longleftrightarrow\quad\lim_{L\to\infty}\frac{|W-W_{\can}|}{L^d}=0.
\end{align}
Equivalently, this condition can be expressed as
\begin{align}
\forall\epsilon>0\quad\exists L_\epsilon\quad    \forall L\ge L_\epsilon,\qquad    \frac{|W-W_{\can}|}{L^d}<\epsilon.
\end{align}
It is therefore sufficient to show that the error defined in Eq.~\eqref{13-1} can be made arbitrarily small per unit volume.
Equation~\eqref{bb} implies that
\begin{align}
    \frac{{\rm Eq.~(\ref{13-1})}}{L^d} \le  (1+2l)^d e^{-\kappa[l-r_0-vt]}.
\end{align}
For $t=\mathcal{O}(L^0)$ and any $\epsilon>0$, one can choose $l_\epsilon$ independent of $L$ such that
\begin{align}
    (1+2l_\epsilon)^d e^{-\kappa[l_\epsilon-r_0-vt]}<\epsilon.
\end{align}
Consequently,
\begin{align}
    \forall\epsilon>0\quad \exists l_\epsilon\quad \exists L_\epsilon\quad \forall L\ge L_\epsilon,\qquad \frac{{\rm Eq.~(\ref{13-1})}}{L^d}<\epsilon.
\end{align}
Since $\epsilon>0$ is arbitrary, the error in Eq.~\eqref{13-1} is subextensive. 
Hence, Eq.~\eqref{feed-eq} remains valid for $s=0$.
Combining the two cases, we find that Eq.~\eqref{feed-eq} holds when
\begin{equation}
    l=\mathcal{O}(L^s),\qquad
        t=\mathcal{O}(L^s),\qquad 0\le s<1,
    \label{ll-summary}
\end{equation}
where, for $s>0$, the auxiliary length $l$ is
chosen such that $l-r_0-vt\ge cL^s$ for some
constant $c>0$. For $s=0$, $l$ is chosen
sufficiently large but independently of $L$,
as explained above.

\subsection{Measurement processes~(Fig. \ref{speading}(b))}

\begin{figure}[tp]
 \begin{center}
  \includegraphics[scale=0.1]{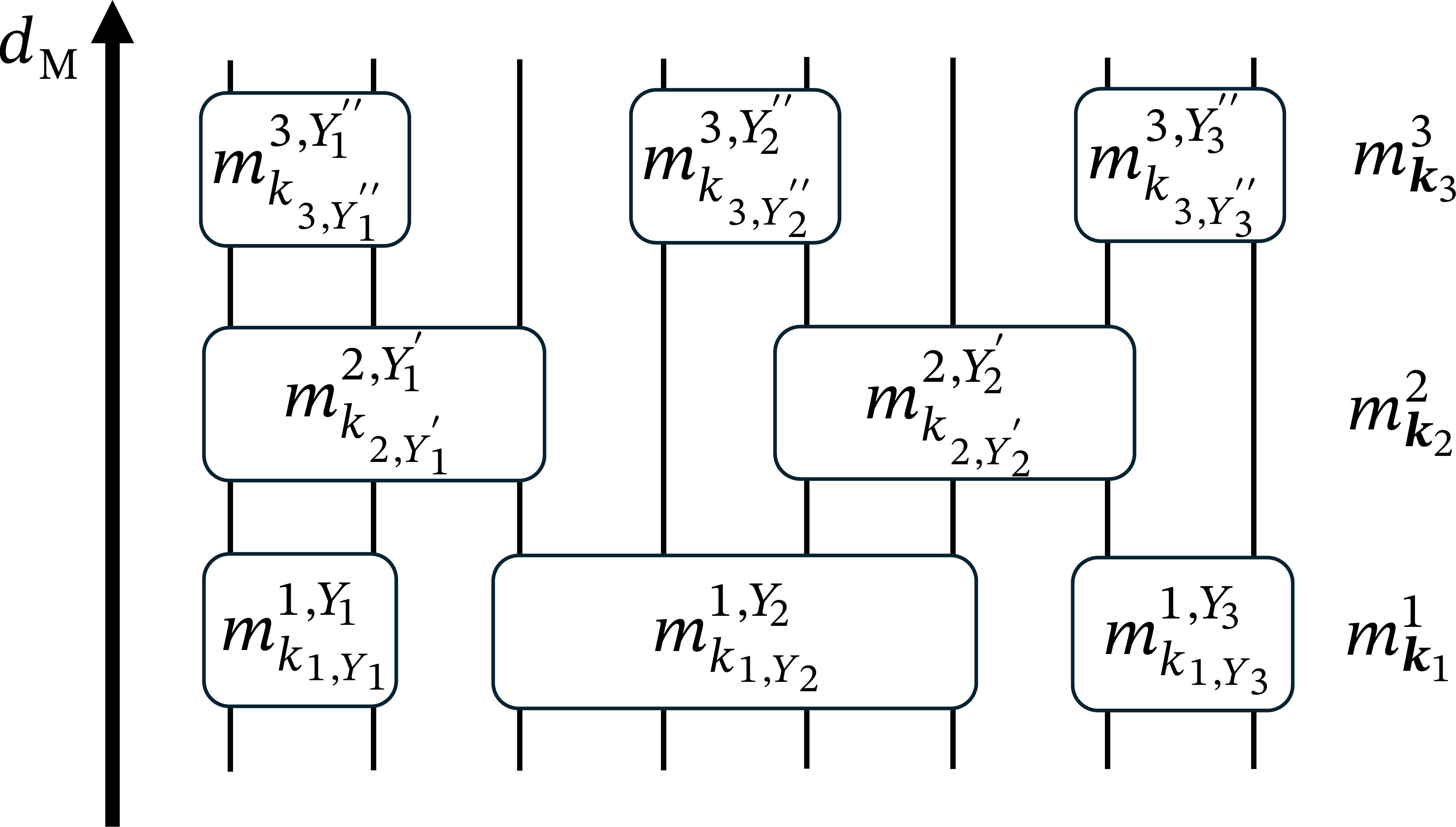}
  \caption{Schematic illustration of a local measurement circuit of depth $d_{\mathrm M}=3$ in a one-dimensional system.
  The measurement operator is $M_{\bm k}=m_{\bm k_3}^{3}m_{\bm k_2}^{2}m_{\bm k_1}^{1}$, with $m_{\bm k_1}^{1}=m_{k_{1,Y_1}}^{1,Y_1}m_{k_{1,Y_2}}^{1,Y_2}m_{k_{1,Y_3}}^{1,Y_3}$, $m_{\bm k_2}^{2}=m_{k_{2,Y_1^{\prime}}}^{2,Y_1^{\prime}}m_{k_{2,Y_2^{\prime}}}^{2,Y_2^{\prime}}$, and $m_{\bm k_3}^{3}=m_{k_{3,Y_1^{\prime\prime}}}^{3,Y_1^{\prime\prime}}m_{k_{3,Y_2^{\prime\prime}}}^{3,Y_2^{\prime\prime}}m_{k_{3,Y_3^{\prime\prime}}}^{3,Y_3^{\prime\prime}}$.
  The diameters of the corresponding regions are $(1,3,1)$, $(2,2)$, and $(1,1,1)$ for $(Y_1,Y_2,Y_3)$, $(Y_1^{\prime},Y_2^{\prime})$, and $(Y_1^{\prime\prime},Y_2^{\prime\prime},Y_3^{\prime\prime})$, respectively.
 \label{m-ope}}
  \end{center}
\end{figure}

Next, we consider
\begin{align}
    \sum_{\bm{k}}\sum_{X}{\rm Tr}[\rho_{i}^{\rm MITE}U_1^{\dagger}M_{\bm k}^{\dagger}h_{{\bm k},X}^{[l]}(t)M_{\bm k}U_1].\label{18-1}
\end{align}
We first set
\begin{align}
    G_X^{[l]}(t):=\sum_{\bm{k}}M_{\bm k}^{\dagger}h_{{\bm k},X}^{[l]}(t)M_{\bm k}.
    \label{60}
\end{align}
We now clarify the condition imposed on the measurement operator $M_{\bm k}$.
We represent the measurement operator corresponding to outcome $\bm k$ by a depth-$d_{\mathrm M}$ local measurement circuit (Fig.~\ref{m-ope}):
\begin{align}
 M_{\bm k}&=m_{\bm k_{d_{\mathrm M}}}^{d_{\mathrm M}}m_{\bm k_{d_{\mathrm M}-1}}^{d_{\mathrm M}-1}\cdots m_{\bm k_1}^{1},
 \label{eq:measurement-depth-sup}\\
 m_{\bm k_n}^{n}&=\bigotimes_{Y\in\mathcal{F}_n}m_{k_{n,Y}}^{n,Y}.
 \label{eq:measurement-layer-sup}
\end{align}
Here, $m_{k_{n,Y}}^{n,Y}$ is supported on the region $Y$, as explained in the main paper.
The local completeness relation is
\begin{align*}
    \sum_{k_{n,Y}}
    \bigl(m_{k_{n,Y}}^{n,Y}\bigr)^\dagger m_{k_{n,Y}}^{n,Y}
    =\id_Y.
\end{align*}
Each $\mathcal{F}_n$ is a collection of mutually disjoint local regions $Y$.
We define the diameter of the region $Y$ by
\begin{align}
    {\rm diam}(Y):=\max\{{\rm dist}(y_1,y_2):y_1,y_2\in Y\},\label{dia-sup}
\end{align}
where ${\rm dist}(y_1,y_2)$ is the Euclidean distance, and
\begin{align}
    \mathscr{D}:=\max_{n}\max_{Y\in\mathcal{F}_n}{\rm diam}(Y)+1.
\end{align}

Let ${\rm supp}(G_X^{[l]}(t))$ be a region on which $G_X^{[l]}(t)$ is supported.
By definition, the truncated Hamiltonian $H_{\bm k}^{X^l}$ contains only the
Hamiltonian terms $\tilde h_{\bm k^{(Z)},Z}$ with
$Z\subseteq X^l$.
The term $\tilde h_{\bm k^{(Z)},Z}$ may depend on the
measurement outcome $k_{n,Y}$ only if there is a site
$y_0\in Y$ such that
${\rm dist}(Z,\{y_0\})\le\ell$.
Such a site $y_0$ lies in $X^{l+\ell}$.
Thus, $h_{\bm k,X}^{[l]}(t)$ is independent of $k_{n,Y}$
whenever $Y$ is disjoint from $X^{l+\ell}$.
When summing over the outcomes, we must account for
both the support of $h_{\bm k,X}^{[l]}(t)$ and its
dependence on those outcomes.
For measurement regions disjoint from $X^{l+\ell}$,
this operator is independent of the corresponding
outcomes and commutes with the measurement operators.
In the first layer processed, measurement regions
disjoint from $X^{l+\ell}$ can therefore be removed
by local completeness.
Consider the measurement regions in this layer that
overlap $X^{l+\ell}$.
After summing over the outcomes of this layer,
the resulting operator is supported within the union
of these regions and $X^{l+\ell}$.
In the next layer, measurement regions disjoint from
the enlarged region can be removed in the
same way.
Only regions overlapping the enlarged region can extend
it further, and we repeat this procedure for the
remaining layers.

Within each layer, the measurement regions are mutually
disjoint and have diameter at most $\mathscr D$.
Each layer can therefore increase the distance from $X$
by at most $\mathscr D-1$.
After $d_{\mathrm M}$ layers, every site in ${\rm supp}(G_X^{[l]}(t))$
is contained in $X^{R_M}$, where
\begin{align}
    R_M:=l+\ell+[\mathscr D-1]d_{\mathrm M}
    \label{RM}
\end{align}
is the radius of $X^{R_M}$.
That is,
\begin{align}
    {\rm supp}(G_X^{[l]}(t))\subseteq X^{R_M}.
\end{align}
In particular, the diameter of ${\rm supp}(G_X^{[l]}(t))$ is bounded as
\begin{align}
\label{65}
    {\rm diam}\left({\rm supp}(G_X^{[l]}(t))\right)\le 2R_M
    =
    2(l+\ell)+2\left[\mathscr{D}-1\right]d_{\mathrm M}.
\end{align}
For consistency with the extended MITE condition discussed below, we assume that $M_{\bm k}$ satisfies
\begin{align}
\left[\mathscr{D}-1\right]d_{\mathrm M}
    =
    \mathcal{O}(L^{s}), \qquad 0\le s<1.
    \label{measure}
\end{align}
In other words, we assume that the measurement operator is a tensor product of low-depth and local measurements. 
Under this assumption, $R_M=\mathcal{O}(L^s)$, so $G_X^{[l]}(t)$ acts only on an $\mathcal{O}(L^s)$-local region.

\subsection{Premeasurement processes~(Fig. \ref{speading}(c))}

With the definition in Eq.~(\ref{60}),
\begin{align}
    {\rm Eq.~(\ref{18-1})}=\sum_{X}{\rm Tr}[\rho_{i}^{\rm MITE}U_1^{\dagger}G_X^{[l]}(t)U_1].
\end{align}
The unitary operator $U_1$, unlike the feedback unitary operator, is independent of the measurement outcome and can be written as
\begin{align}
    U_{1}:=U_{1}(t^{\prime})=\Torder\exp\left(-i\int_0^{t^{\prime}}H_1(u)du\right),
\end{align}
where $t^{\prime}$ denotes the duration for which the unitary operator $U_1$ is applied.
As we mentioned before, we assume that $H_1(u)=H_1$ is time-independent, although the generalization to time-dependent cases is straightforward.
We choose $l^{\prime}$ such that
\begin{align}
    R_M\le l^{\prime}.
    \label{3l}
\end{align}
We now define an operator supported on $X^{l^{\prime}}$ by
\begin{align}
    G_{X}^{[l^{\prime},l]}(t,t^{\prime})
    &:=
    \frac{\bm 1_{\overline{X^{l^{\prime}}}}}{{\rm Tr}\bm 1_{\overline{X^{l^{\prime}}}}}
    \otimes
    {\rm Tr}_{\overline{X^{l^{\prime}}}}
    [U_1^{\dagger}G_{X}^{[l]}(t)U_1].
    \label{30-2}
\end{align}
In this case,
\begin{align}
    &\left|\sum_{X}{\rm Tr}[\rho_{i}^{\rm MITE}U_1^{\dagger}G_X^{[l]}(t)U_1]
    -\sum_{X}{\rm Tr}[\rho_{i}^{\rm MITE}G_{X}^{[l^{\prime},l]}(t,t^{\prime})]\right|\label{20-1}\notag\\
    &=\left|\sum_{X}{\rm Tr}[\rho_{i}^{\rm MITE}(U_1^{\dagger}G_X^{[l]}(t)U_1-G_{X}^{[l^{\prime},l]}(t,t^{\prime}))]\right|\notag\\   
    &\le\sum_{X}
    \|U_1^{\dagger}G_X^{[l]}(t)U_1-G_{X}^{[l^{\prime},l]}(t,t^{\prime})\|_{\infty}.
\end{align}
Applying the Lieb--Robinson bound~\cite{Lieb1972,Bravyi2006,gong2022bounds} to the additional spreading generated by $U_1$, and using the fact that $G_X^{[l]}(t)$ is already contained in $X^{R_M}$, we obtain
\begin{align}
    {\rm Eq.~(\ref{20-1})}
    \le
    \sum_{X}c_2^{\prime}
    e^{-\kappa^{\prime}(l^{\prime}-R_M-v^{\prime}t^{\prime})},
    \label{20-2}
\end{align}
where
\begin{align}
    c_2^{\prime}
    =
    c_2
    \|U_1^{\dagger}G_{X}^{[l]}(t)U_1\|_{\infty}
    |X^{l^{\prime}}|.
    \label{21-11}
\end{align}
Here, $c_2$ is a constant independent of the system size, 
$v'$ is the Lieb-Robinson velocity associated with $H_1$, $\kappa^{\prime}>0$ is the spatial decay rate of the Lieb--Robinson tail,
and $|X^{l^{\prime}}|$ denotes the number of sites in the region $X^{l^{\prime}}$.
In spatial dimension $d$, we can estimate $|X^{l^{\prime}}|\sim (1+2l^{\prime})^d$.
Moreover,
\begin{align}
    \|U_1^{\dagger}G_{X}^{[l]}(t)U_1\|_{\infty}
    &=\|G_{X}^{[l]}(t)\|_{\infty}\notag\\
    &=\|\sum_{\bm{k}}M_{\bm k}^{\dagger}h_{{\bm k},X}^{[l]}(t)M_{\bm k}\|_{\infty}.
    \label{21-1}
\end{align}
Since $M_{\bm k}^{\dagger}h_{{\bm k},X}^{[l]}(t)M_{\bm k}$ is Hermitian,
\begin{align}
    &{\rm Eq.~(\ref{21-1})}\notag\\
    &=\sup_{\|\ket{\psi}\|=1}\left|\bra{\psi}\sum_{\bm k}M_{\bm k}^{\dagger}h_{{\bm k},X}^{[l]}(t)M_{\bm k}\ket{\psi}\right|,
    \label{22-1}
\end{align}
where $\ket{\psi}$ is a unit vector.
This expression can be further transformed as
\begin{align}
    &{\rm Eq.~(\ref{22-1})}\notag\\
    &\leq\sup_{\|\ket{\psi}\|=1}\sum_{\bm k}\|M_{\bm k}\ket{\psi}\|^2
    \left|
    \frac{\bra{\psi}M_{\bm k}^{\dagger}}{\|M_{\bm k}\ket{\psi}\|}
    h_{{\bm k},X}^{[l]}(t)
    \frac{M_{\bm k}\ket{\psi}}{\|M_{\bm k}\ket{\psi}\|}
    \right|\notag\\
    &\le
    \sup_{\|\ket{\psi}\|=1}\sum_{\bm k}\|M_{\bm k}\ket{\psi}\|^2
    \max_{\bm k}
    \left|
    \frac{\bra{\psi}M_{\bm k}^{\dagger}}{\|M_{\bm k}\ket{\psi}\|}
    h_{{\bm k},X}^{[l]}(t)
    \frac{M_{\bm k}\ket{\psi}}{\|M_{\bm k}\ket{\psi}\|}
    \right|\notag\\
    &=
    \sup_{\|\ket{\psi}\|=1}
    \max_{\bm k}
    \left|
    \frac{\bra{\psi}M_{\bm k}^{\dagger}}{\|M_{\bm k}\ket{\psi}\|}
    h_{{\bm k},X}^{[l]}(t)
    \frac{M_{\bm k}\ket{\psi}}{\|M_{\bm k}\ket{\psi}\|}
    \right|\notag\\
    &\leq
    \max_{\bm k}\|h_{{\bm k},X}^{[l]}(t)\|_{\infty}.
\end{align}
Thus,
\begin{align}
    \|U_1^{\dagger}G_{X}^{[l]}(t)U_1\|_{\infty}
    \le
    \max_{\bm k}\|h_{{\bm k},X}^{[l]}(t)\|_{\infty}.
    \label{23-1}
\end{align}
By the triangle inequality,
\begin{align}
    &{\rm Eq.~(\ref{23-1})}\notag\\
    &\le
    \max_{\bm k}
    \left[
    \|h_{{\bm k},X}(t)\|_{\infty}
    +
    \|h_{{\bm k},X}^{[l]}(t)-h_{{\bm k},X}(t)\|_{\infty}
    \right]\notag\\
    &=
    \max_{\bm k}
    \left[
    \|U_{\bm k}^{\dagger}(t)h_XU_{\bm k}(t)\|_{\infty}
    +
    \|h_{{\bm k},X}(t)-h_{{\bm k},X}^{[l]}(t)\|_{\infty}
    \right]\notag\\
    &=
    \max_{\bm k}
    \left[
    \|h_X\|_{\infty}
    +
    \|h_{{\bm k},X}(t)-h_{{\bm k},X}^{[l]}(t)\|_{\infty}
    \right].
    \label{23-2}
\end{align}
As in the derivation of Eq.~(\ref{15-2}), the Lieb--Robinson bound gives
\begin{align}
    {\rm Eq.~(\ref{23-2})}
    \le
    \|h_X\|_{\infty}
    +
    C\|h_X\|_{\infty}(1+2l)^d e^{-\kappa(l-r_0-vt)}.
\end{align}
Thus, when $l$ satisfies Eq.~\eqref{ll-summary}, where $t$ is the duration of the feedback control, the second term vanishes in the thermodynamic limit.
Consequently,
\begin{align}
    \|U_1^{\dagger}G_{X}^{[l]}(t)U_1\|_{\infty}
    \lesssim
    \|h_X\|_{\infty}.
    \label{21-10}
\end{align}
Therefore, from Eqs.~(\ref{20-2}), (\ref{21-11}), and (\ref{21-10}),
\begin{align}
    {\rm Eq.~(\ref{20-1})}
    &\le
    \sum_{X}c_2
    \|U_1^{\dagger}G_{X}^{[l]}(t)U_1\|_{\infty}
    |X^{l^{\prime}}|
    e^{-\kappa^{\prime}(l^{\prime}-R_M-v^{\prime}t^{\prime})}\notag\\
    &\lesssim
    \sum_{X}c_2
    \|h_X\|_{\infty}
    (1+2l^{\prime})^d
    e^{-\kappa^{\prime}(l^{\prime}-R_M-v^{\prime}t^{\prime})}.
\end{align}
Now, 
as well as Eq.~\eqref{ll-summary},
we assume for $l^{\prime}$ that
\begin{equation}
    l^{\prime}=\mathcal{O}(L^s),\qquad
    R_M\le l^{\prime},\qquad
        t^{\prime}=\mathcal{O}(L^s),\qquad 0\le s<1,
    \label{llll}
\end{equation}
where, for $s>0$, the auxiliary length $l'$ is
chosen such that $l'-R_M-v't'\ge c'L^s$ for some
constant $c'>0$. For $s=0$, $l'$ is chosen
sufficiently large but independently of $L$,
in a similar manner to Eq.~\eqref{ll-summary}.
Then, in the thermodynamic limit
the operator $U_1^{\dagger}G_{X}^{[l]}(t)U_1$ in the first term of Eq.~(\ref{20-1}) can be approximated by $G_{X}^{[l^{\prime},l]}(t,t^{\prime})$:
\begin{align}
    \sum_{X}{\rm Tr}[\rho_{i}^{\rm MITE}U_1^{\dagger}G_X^{[l]}(t)U_1]
    \simeq
    \sum_{X}{\rm Tr}[\rho_{i}^{\rm MITE}G_{X}^{[l^{\prime},l]}(t,t^{\prime})].
\end{align}

We next consider
\begin{align}
    \sum_{X}{\rm Tr}[\rho_{i}^{\rm MITE}G_{X}^{[l^{\prime},l]}(t,t^{\prime})].
\end{align}
Because $\rho_{i}^{\rm MITE}$ is an MITE, and from Eq.~\eqref{llll}, the support radius $l^{\prime}$ of $G_{X}^{[l^{\prime},l]}(t,t^{\prime})$ satisfies
\begin{align}
    l^{\prime}=\mathcal{O}(L^{s})~~(0\le s<1),
    \label{lll}
\end{align}
so, in the thermodynamic limit,
\begin{align}
    \sum_{X}{\rm Tr}[\rho_{i}^{\rm MITE}G_{X}^{[l^{\prime},l]}(t,t^{\prime})]
    \simeq
    \sum_{X}{\rm Tr}[\rho_i^{\rm ens}G_{X}^{[l^{\prime},l]}(t,t^{\prime})].
\end{align}
We now summarize the conditions on $Z$, $l$, $R_M$, and $l^{\prime}$ that have appeared so far.
From Eqs.~(\ref{feedcon}), (\ref{ll-summary}), (\ref{RM}), (\ref{measure}), (\ref{3l}), and (\ref{llll}),
\begin{align}
&{\rm dist}(Z,\{y_0\})\le\ell,
    \qquad \ell=\mathcal O(L^s),\\
    &l=\mathcal{O}(L^s),\qquad
    t=\mathcal{O}(L^s),\\        &R_M=l+\ell+\left[\mathscr{D}-1\right]d_{\mathrm M}
        =\mathcal{O}(L^{s}),\\
        &l^{\prime}=\mathcal{O}(L^{s}),\qquad
    R_M\le l^{\prime},\qquad
        t^{\prime}=\mathcal{O}(L^s)
\end{align}
for $0\leq s<1$.
These conditions mean that the feedback light cone, the measurement causal cone, and the additional light cone generated by $U_1$ all remain within sublinear regions, when we consider the operator spreading of $h_X$.
Note that $l$ and $l'$ are auxiliary scales introduced for the proof, while $\ell$, $t$, $t'$, $\mathscr{D}$, and $d_\mathrm{M}$ are the parameters given in the setup.

If these conditions are satisfied, $U_1^{\dagger}M_{\bm k}^{\dagger}U_{\bm k}(t)^{\dagger}h_{X}U_{\bm k}(t)M_{\bm k}U_1$ can be safely approximated as an $\mathcal{O}(L^s)$-local observable, and
\begin{align}
    {\rm Tr}[H\rho_f]
    &=
    \sum_{\bm{k}}\sum_{X}
    {\rm Tr}[\rho_{i}^{\MITE}
    U_1^{\dagger}M_{\bm k}^{\dagger}U_{\bm k}(t)^{\dagger}h_{X}U_{\bm k}(t)M_{\bm k}U_1]
    \label{31-0}\\
    &\simeq
    \sum_{X}{\rm Tr}[\rho_i^{\rm ens}G_{X}^{[l^{\prime},l]}(t,t^{\prime})]
    \label{31-1}
\end{align}
can be used as an approximation.
If $\rho_i^{\MITE}$ in Eq.~(\ref{31-0}) is replaced by $\rho_i^{\rm ens}$, the same argument leading to Eq.~(\ref{31-1}) applies.
Hence,
\begin{align}
    &\sum_{\bm{k}}\sum_{X}
    {\rm Tr}[\rho_{i}^{\rm ens}
    U_1^{\dagger}M_{\bm k}^{\dagger}U_{\bm k}(t)^{\dagger}h_{X}U_{\bm k}(t)M_{\bm k}U_1]\notag\\
    &\simeq
    \sum_{X}{\rm Tr}[\rho_i^{\rm ens}G_{X}^{[l^{\prime},l]}(t,t^{\prime})].
    \label{31-2}
\end{align}
From Eqs.~(\ref{31-1}) and (\ref{31-2}),
\begin{align}
    {\rm Tr}[H\rho_f]
    \simeq
    \sum_{\bm{k}}\sum_{X}
    {\rm Tr}[\rho_{i}^{\rm ens}
    U_1^{\dagger}M_{\bm k}^{\dagger}U_{\bm k}(t)^{\dagger}h_{X}U_{\bm k}(t)M_{\bm k}U_1]
\end{align}
holds.
Therefore, Eq.~(\ref{10-1}) becomes
\begin{align}
    W^{\MITE}
    &=
    {\rm Tr}[H\rho_{i}^{\MITE}]
    -
    {\rm Tr}[H\rho_f]\notag\\
    &\simeq
    {\rm Tr}[H\rho_{i}^{\rm ens}]
    -
    \sum_{\bm{k}}
    {\rm Tr}[\rho_{i}^{\rm ens}
    U_1^{\dagger}M_{\bm k}^{\dagger}U_{\bm k}(t)^{\dagger}HU_{\bm k}(t)M_{\bm k}U_1].
\end{align}
This coincides with the work extractable from the canonical state, and therefore proves Eq.~\eqref{eq:main-equivalence}.
We summarize the preceding discussion as a theorem.

\begin{theo*}{Extractable work from MITE}
Consider a closed quantum many-body system.
Suppose that the Hamiltonian $H$ is a sum of local observables, $H=\sum_X h_X$, and that the initial state is MITE for $\mathcal{O}(L^s)$-local observables ($0\leq s <1$).
We assume that
the measurement operator defined in Eq.~\eqref{eq:measurement-depth} is a tensor product of low-depth local measurements with $(\mathscr{D}-1)d_\mathrm{M}=\mathcal{O}(L^s)$, and
that the duration of feedback control $t$ and the duration $t^{\prime}$
of the premeasurement unitary are $\mathcal{O}(L^s)$.
In addition, each local feedback term $\tilde h_{\bm k^{(Z)},Z}(u)$
may depend on a measurement outcome $k_{n,Y}$ (an element in $\bm k^{(Z)}$) only if
\begin{align*}
    \dist(Z,Y)\le\ell,
    \qquad \ell=\mathcal O(L^s).
\end{align*}
Then, the work $W$ extractable by measurement and feedback control is asymptotically equal to the work $W_{\can}$ extractable by applying the same protocol when the initial state is the canonical state, and it can be written in terms of information quantities as
\begin{align}
    \beta W\simeq\beta W_{\can}
    =
    I_{\rm QC}-\chi-D(\rho_f^\can\Vert\rho_i^{\can}).
\end{align}
The quantities $I_{\rm QC}$ and $\chi$ are defined in Eqs.~\eqref{eq:IQC} and  \eqref{eq:Holevo}, respectively.
\end{theo*}

\section{S2. Details for the $V$-qubit system}

We provide the details of work extraction from the $V$-qubit system by measurement and feedback control, as discussed in the main text.
As stated in the main text, we consider a cycle and take the Hamiltonian as
\begin{align}
    H=g\sum_{j=1}^{V}\sigma_j^3.
\end{align}
Here, $\sigma_j^3$ denotes the operator in which $\sigma^3$ acts only on the $j$th site:
\begin{align}
\sigma_j^3=\sigma^0\otimes\sigma^0\otimes\cdots\otimes\sigma^3\otimes\sigma^0\otimes\cdots\otimes\sigma^0.
\end{align}

Note that there is no interaction between sites.
We take the initial state to be the canonical state:
\begin{align}
    \rho_i^{\can}
=\frac{e^{-\beta H}}{Z}
=\begin{pmatrix}
    \frac{e^{-g\beta}}{2\cosh(g\beta)} & 0\\
    0 & \frac{e^{g\beta}}{2\cosh(g\beta)}
\end{pmatrix}^{\otimes V}.
\end{align}

We apply the operators $U_1$, $M_{\bm k}$, and $U_{\bm k}$ to the initial state $\rho_i^{\can}$. The final state is then
\begin{align}
    \rho_f^\can=\sum_{\bm k} U_{\bm k} M_{\bm  k} U_1 \rho_i^{\can} U_1^{\dagger}M_{\bm k}^{\dagger}U_{\bm k}^{\dagger}\label{1000-1}.
\end{align}
For simplicity, we take the premeasurement unitary operator to be $U_1=\sigma^0\otimes\sigma^0\otimes\cdots\otimes\sigma^0$.

We describe the measurement operator by the following Kraus operators $M_{\pm,\pm,\cdots,\pm}$ and assume that it is written as a tensor product of the measurement operators $m_{k_l}(\eta)$ on the individual qubits:
\begin{align}
    M_{\pm,\pm,\cdots,\pm}=\bigotimes_{l=1}^{V}m_{k_l}(\eta).
\end{align}
Here, $k_l$ corresponds to the measurement outcome of $\sigma^3$ at site $l$ and can take the value $+$ or $-$. Also,
\begin{align}
    m_{k_l}(\eta)=\frac{1}{\sqrt{2(1+\eta^2)}}\left(\sigma^0\pm\eta\sigma^3\right)~~~(0\le\eta\le1).
\end{align}
This quantum measurement contains an error when $\eta<1$, and $+$ or $-$ denotes the measurement outcome of $\sigma^3$ in the presence of this error \cite{Mochizuki-Hamazaki-MIST}. We take the corresponding feedback operator to be
\begin{align}
    U_{\pm,\pm,\cdots,\pm}=\bigotimes_{l=1}^{V}u_{k_l},
\end{align}
where $u_{k_l}$ is a unitary operator acting on site $l$. The explicit forms of $u_+$ and $u_-$ will be given below.
The definition of $k_l$ is as given above.

Next, consider the one-qubit density operator
\begin{align}
    \rho_i^{1}:=\frac{e^{-\beta g\sigma^3}}{2\cosh(g\beta)}=\frac{1}{2\cosh(g\beta)}
    \begin{pmatrix}
        e^{-g\beta}&0\\
        0&e^{g \beta}
    \end{pmatrix}.
\end{align}
We use this state in the following.
We first consider the case in which the measurement operator $m_-$ is applied.
Then,
\begin{align}
    m_{-}\rho_{i}^1m_{-}^{\dagger}=\frac{1}{4(1+\eta^2)\cosh(g\beta)}\left[(1-\eta)^2e^{-g\beta}\ket{\uparrow}\bra{\uparrow}+(1+\eta)^2e^{g\beta}\ket{\downarrow}\bra{\downarrow}\right].
\end{align}
Next, we determine the feedback unitary operator $u_-$ that minimizes
\begin{align}
    {\rm Tr}[u_-m_-\rho_i^1 m_-^{\dagger}u_-^{\dagger}g\sigma^3]
\end{align}
with respect to $u_-$.
For this purpose, we should make the probability of obtaining $\ket{\downarrow}\bra{\downarrow}$ larger than that of obtaining $\ket{\uparrow}\bra{\uparrow}$. Therefore, the feedback operator $u_-$ to be constructed may be chosen as
\begin{align}
    u_-&=c\sigma^0
\end{align}
with $|c|^2=1$. Under the feedback operation implemented in this way,
\begin{align}
    u_-m_{-}\rho_{i}^1m_{-}^{\dagger}u_-^{\dagger}=\frac{1}{4(1+\eta^2)\cosh(g\beta)}\left[(1-\eta)^2e^{-g\beta}\ket{\uparrow}\bra{\uparrow}+(1+\eta)^2e^{g\beta}\ket{\downarrow}\bra{\downarrow}\right]\label{123-2}
\end{align}
is obtained.

Next, consider the case in which the measurement operator $m_+$ is applied. Then,
\begin{align}
    m_{+}\rho_{i}^{1}m_{+}^{\dagger}=\frac{1}{4(1+\eta^2)\cosh(g\beta)}\left[(1+\eta)^2e^{-g\beta}\ket{\uparrow}\bra{\uparrow}+(1-\eta)^2e^{g\beta}\ket{\downarrow}\bra{\downarrow}\right].
\end{align}
Therefore, the optimal feedback operator to be constructed depends on the following cases.
\vskip\baselineskip
\underline{$\rm(\hspace{.18em}i\hspace{.18em})$ Case $(1+\eta)^2e^{-g\beta}\ge (1-\eta)^2e^{g\beta}$ (high temperature or high measurement accuracy)}
\vskip\baselineskip
In this case, the feedback unitary operator is
\begin{align}
    u_+=c\left(\ket{\downarrow}\bra{\uparrow}+\ket{\uparrow}\bra{\downarrow}\right)=c\sigma^1,
\end{align}
where $|c|^2=1$.
This feedback unitary operator can be realized by applying a magnetic field, as explained below.
If we set
\begin{align}
    H_+=h^{\prime}\sigma^1,
\end{align}
then
\begin{align}
    u_{+}=\exp\left(-i\int_0^{t}H_+ dt^{\prime}\right)=\cos(h^{\prime} t)\sigma^0-i\sin(h^{\prime} t)\sigma^1.
\end{align}
Thus, if the external field $h^{\prime}$ is adjusted according to the feedback-control time $t$ so as to satisfy
\begin{align}
    \cos(h^{\prime}t)=0,\quad\sin(h^{\prime}t)=\mp 1,\label{feedtime}
\end{align}
then the desired unitary operator is obtained, with $c=\pm i$. Under the feedback operation implemented in this way,
\begin{align}
    u_+m_{+}\rho^1_i m_{+}^{\dagger}u_+^{\dagger}=\frac{1}{4(1+\eta^2)\cosh(g\beta)}\left[(1+\eta)^2e^{-g\beta}\ket{\downarrow}\bra{\downarrow}+(1-\eta)^2e^{g\beta}\ket{\uparrow}\bra{\uparrow}\right].\label{123-1}
\end{align}
Therefore,
\begin{align}
\rho_f^1&:=u_+ m_+ u_1 \rho_i^1 u_1^{\dagger}m_+^{\dagger}u_+^{\dagger}+u_- m_- u_1 \rho_i^1 u_1^{\dagger}m_-^{\dagger}u_-^{\dagger}\notag\\
&=\frac{1}{2(1+\eta^2)}\left[(1-\eta)^2\ket{\uparrow}\bra{\uparrow}+(1+\eta)^2\ket{\downarrow}\bra{\downarrow}\right]
\end{align}
and hence
\begin{align}
    {\rm Tr}[H_{f}^1\rho_{ f}^1]=\frac{-2\eta g}{1+\eta^2},
\end{align}
where
\begin{align}
    H_f^1:=g\sigma^3.
\end{align}

Since we consider noninteracting $V$ qubits, we have
\begin{align}
   {\rm Tr}[\rho_f H_f]=
    \frac{-2\eta g}{1+\eta^2}V.
    \label{123-5}
\end{align}
Moreover,
\begin{align}
    {\rm Tr}[\rho_i^{\can} H_i]=-gV\tanh(g\beta).
\end{align}
Therefore, the extractable work is
\begin{align}
    W_{\rm opt, loc}={\rm Tr}[H_{\rm i}\rho_{\rm i}]-{\rm Tr}[H_{\rm f}\rho_{\rm f}]=-gV\tanh(g\beta)+\frac{2\eta g}{1+\eta^2}V.
    \label{work1}
\end{align}

Next, we compute the corresponding information quantities, namely, the mutual information, the Holevo information, and the relative entropy. We also compute their sum, which is the right-hand side of Eq.~(\ref{eq:canonical-identity}), and check that it indeed agrees with the extractable work.

First, we compute
$\sum_{{\bm{k}}}P_{\bm{k}} S(\rho_{\bm{k}}^{\prime})$.
Since we consider noninteracting tensor-product structure, we find 
\begin{align}
    P_{{\bm{k}}}={\rm Tr}[M_{{\bm{k}}}M_{{\bm{k}}}^{\dagger}\rho_i^{\can}]:=\prod_{l=1}^{V}p_{k_l}
\end{align}
with
\begin{align}
    p_+={\rm Tr}[m_{+}^{\dagger}m_{+}\rho_i^1]=\frac{(1+\eta)^2 e^{-g\beta}+(1-\eta)^2 e^{g\beta}}{4(1+\eta^2)\cosh(g\beta)},
\end{align}
\begin{align}
    p_-=1-p_{+}=\frac{(1-\eta)^2 e^{-g\beta}+(1+\eta)^2 e^{g\beta}}{4(1+\eta^2)\cosh(g\beta)}.
\end{align}
Also,
\begin{align}
    S(\rho_{{\bm{k}}}^{\prime})=S\left(\frac{U_{\bm{k}} M_{\bm{k}} \rho_i^{\can} M_{\bm{k}}^{\dagger}U_{\bm{k}}^{\dagger}}{P_{{\bm{k}}}}\right)
    =\sum_{l=1}^{V}S\left(\frac{u_{k_l}m_{k_l}\rho_i^{1}m_{k_l}^{\dagger}u_{k_l}^{\dagger}}{p_{k_l}}\right)
    =\sum_{l=1}^{V}S(\rho_{k_l}^1).
\end{align}
Thus,
\begin{align}
    \sum_{{\bm{k}}}P_{\bm{k}} S(\rho_{\bm{k}}^{\prime})=V\left(\sum_{k_1=\pm}p_{k_1}S(\rho_{k_1}^1)\right).
\end{align}

From Eq.~(\ref{123-1}),
\begin{align}
    \rho_+^{1}=\frac{u_+ m_+ u_1 \rho_i^1 u_1^{\dagger}m_+^{\dagger}u_+^{\dagger}}{p_+}=\frac{1}{4p_+(1+\eta^2)\cosh(g\beta)}\left[(1+\eta)^2e^{-g\beta}\ket{\downarrow}\bra{\downarrow}+(1-\eta)^2e^{g\beta}\ket{\uparrow}\bra{\uparrow}\right].
\end{align}

Similarly, from Eq.~(\ref{123-2}),
\begin{align}
    \rho_-^{1}&=\frac{u_- m_- u_1 \rho_i^1 u_1^{\dagger}m_-^{\dagger}u_-^{\dagger}}{p_-}\notag\\
    &=\frac{1}{4p_-(1+\eta^2)\cosh(g\beta)}\left[(1-\eta)^2e^{-g\beta}\ket{\uparrow}\bra{\uparrow}+(1+\eta)^2e^{g\beta}\ket{\downarrow}\bra{\downarrow}\right].
\end{align}
Hence, after some calculation, we obtain
\begin{align}
    &\sum_{{\bm{k}}}P_{\bm{k}} S(\rho_{\bm{k}}^{\prime})=V\left(\sum_{k_1=\pm}p_{k_1}S(\rho_{k_1}^1)\right)\notag\\
    &=V\left(-\frac{(1+\eta)^2e^{-g\beta}}{4(1+\eta^2)\cosh(g\beta)}\log\left[\frac{(1+\eta)^2e^{-g\beta}}{(1+\eta)^2e^{-g \beta}+(1-\eta)^2e^{g \beta}}\right]\right.\left.
    -\frac{(1-\eta)^2e^{-g\beta}}{4(1+\eta^2)\cosh(g\beta)}\log\left[\frac{(1-\eta)^2e^{-g\beta}}{(1-\eta)^2e^{-g \beta}+(1+\eta)^2e^{g \beta}}\right]\right.\nonumber\\
    &\left.
    -\frac{(1-\eta)^2e^{g\beta}}{4(1+\eta^2)\cosh(g\beta)}\log\left[\frac{(1-\eta)^2e^{g\beta}}{(1+\eta)^2e^{-g \beta}+(1-\eta)^2e^{g \beta}}\right]\right.\left.
    -\frac{(1+\eta)^2e^{g\beta}}{4(1+\eta^2)\cosh(g\beta)}\log\left[\frac{(1+\eta)^2e^{g\beta}}{(1-\eta)^2e^{-g \beta}+(1+\eta)^2e^{g \beta}}\right]
    \right).
\end{align}

Next, we compute $S(\rho_i^{\can})$:
\begin{align}
    S(\rho_i^{\can})=VS(\rho_i^1)=V\left(-\frac{e^{-g \beta}}{2\cosh(g\beta)}\log\left[\frac{e^{-g \beta}}{2\cosh(g\beta)}\right]-\frac{e^{g \beta}}{2\cosh(g\beta)}\log\left[\frac{e^{g \beta}}{2\cosh(g\beta)}\right]\right).
\end{align}

We can also compute $S(\rho_f)$, represented as
\begin{align}
    S(\rho_f)=VS\left[\sum_{k_1=\pm}u_{k_1} m_{k_1} \rho_i^1 m_{k_1}^{\dagger}u_{k_1}^{\dagger}\right].
\end{align}
Using Eqs.~(\ref{123-1}) and (\ref{123-2}), we have
\begin{align}
    S(\rho_f)\notag&=VS\left[\frac{1}{4(1+\eta^2)\cosh(g\beta)}\left(2(1+\eta)^2\cosh(g\beta)\ket{\downarrow}\bra{\downarrow}+2(1-\eta)^2\cosh(g\beta)\ket{\uparrow}\bra{\uparrow}\right)\right]\notag\\
    &=V\left[-\frac{(1+\eta)^2}{2(1+\eta^2)}\log\left(\frac{(1+\eta)^2}{2(1+\eta^2)}\right)-\frac{(1-\eta)^2}{2(1+\eta^2)}\log\left(\frac{(1-\eta)^2}{2(1+\eta^2)}\right)\right].
\end{align}

From the above results, the mutual information $I_{\rm QC}$ is
\begin{align}
    I_{\rm QC}&=S(\rho_i^{\can})-\sum_{{\bm{k}}}P_{\bm{k}} S(\rho_{\bm{k}}^{\prime})\notag\\
    &=V\left(-\frac{e^{-g \beta}}{2\cosh(g\beta)}\log\left[\frac{e^{-g \beta}}{2\cosh(g\beta)}\right]-\frac{e^{g \beta}}{2\cosh(g\beta)}\log\left[\frac{e^{g \beta}}{2\cosh(g\beta)}\right]\right)\notag\\
    &-V\left(-\frac{(1+\eta)^2e^{-g\beta}}{4(1+\eta^2)\cosh(g\beta)}\log\left[\frac{(1+\eta)^2e^{-g\beta}}{(1+\eta)^2e^{-g \beta}+(1-\eta)^2e^{g \beta}}\right]\right.\notag\\
    &\left.
    -\frac{(1-\eta)^2e^{-g\beta}}{4(1+\eta^2)\cosh(g\beta)}\log\left[\frac{(1-\eta)^2e^{-g\beta}}{(1-\eta)^2e^{-g \beta}+(1+\eta)^2e^{g \beta}}\right]\right.\notag\\
    &\left.
    -\frac{(1-\eta)^2e^{g\beta}}{4(1+\eta^2)\cosh(g\beta)}\log\left[\frac{(1-\eta)^2e^{g\beta}}{(1+\eta)^2e^{-g \beta}+(1-\eta)^2e^{g \beta}}\right]\right.\notag\\
        &\left.
    -\frac{(1+\eta)^2e^{g\beta}}{4(1+\eta^2)\cosh(g\beta)}\log\left[\frac{(1+\eta)^2e^{g\beta}}{(1-\eta)^2e^{-g \beta}+(1+\eta)^2e^{g \beta}}\right]
    \right).
\end{align}
For example, for $\eta=1$ (an error-free measurement), the mutual information $I_{\rm QC}$ becomes equal to the von Neumann entropy $S(\rho_i)$ of the initial state.
For $\eta=0$ (a maximally erroneous measurement), the mutual information $I_{\rm QC}$ becomes
vanishing.

Next, we compute the Holevo information.
\begin{align}
    \chi&=S(\rho_f)-\sum_{{\bm{k}}}P_{\bm{k}} S(\rho_{\bm{k}}^{\prime})\notag\\
    &=V\left[-\frac{(1+\eta)^2}{2(1+\eta^2)}\log\left(\frac{(1+\eta)^2}{2(1+\eta^2)}\right)-\frac{(1-\eta)^2}{2(1+\eta^2)}\log\left(\frac{(1-\eta)^2}{2(1+\eta^2)}\right)\right]\notag\\
    &-V\left(-\frac{(1+\eta)^2e^{-g\beta}}{4(1+\eta^2)\cosh(g\beta)}\log\left[\frac{(1+\eta)^2e^{-g\beta}}{(1+\eta)^2e^{-g \beta}+(1-\eta)^2e^{g \beta}}\right]\right.\notag\\
    &\left.
    -\frac{(1-\eta)^2e^{-g\beta}}{4(1+\eta^2)\cosh(g\beta)}\log\left[\frac{(1-\eta)^2e^{-g\beta}}{(1-\eta)^2e^{-g \beta}+(1+\eta)^2e^{g \beta}}\right]\right.\notag\\
    &\left.
    -\frac{(1-\eta)^2e^{g\beta}}{4(1+\eta^2)\cosh(g\beta)}\log\left[\frac{(1-\eta)^2e^{g\beta}}{(1+\eta)^2e^{-g \beta}+(1-\eta)^2e^{g \beta}}\right]\right.\notag\\
        &\left.
    -\frac{(1+\eta)^2e^{g\beta}}{4(1+\eta^2)\cosh(g\beta)}\log\left[\frac{(1+\eta)^2e^{g\beta}}{(1-\eta)^2e^{-g \beta}+(1+\eta)^2e^{g \beta}}\right]
    \right).
\end{align}
For example, for the infinite-temperature state ($\beta=0$), the Holevo information becomes
zero.

Next, we compute the relative entropy. From Eq.~(\ref{123-5}),
\begin{align}
    D(\rho_f^\can\Vert\rho_i^{\can})&=-S(\rho_f)+\beta{\rm Tr}[H_f \rho_f]+\log Z\notag\\
    &=-V\left[-\frac{(1+\eta)^2}{2(1+\eta^2)}\log\left(\frac{(1+\eta)^2}{2(1+\eta^2)}\right)-\frac{(1-\eta)^2}{2(1+\eta^2)}\log\left(\frac{(1-\eta)^2}{2(1+\eta^2)}\right)\right]+\beta\frac{-2\eta g}{1+\eta^2}V+V\log(2\cosh(g\beta)).
\end{align}
For example, for an infinite-temperature state with an error-free measurement ($\beta=0, \eta=1$), the relative entropy becomes
\begin{align}
    D(\rho_f^\can\Vert\rho_i^{\can})
    &=V\log2.
\end{align}

Finally, we compute the sum of the information quantities, i.e., the right-hand side of Eq.~(\ref{eq:canonical-identity}), and check that it indeed agrees with the extractable work.
\begin{align}
    &I_{\rm QC}-\chi-D(\rho_f^\can\Vert\rho_i^{\can})\notag\\
    &=V\left(-\frac{e^{-g \beta}}{2\cosh(g\beta)}\log\left[\frac{e^{-g \beta}}{2\cosh(g\beta)}\right]-\frac{e^{g \beta}}{2\cosh(g\beta)}\log\left[\frac{e^{g \beta}}{2\cosh(g\beta)}\right]\right)-\beta\frac{-2\eta g}{1+\eta^2}V-V\log(2\cosh(g\beta))\notag\\
    &=V\left(-g\beta\tanh(g\beta)+\log(2\cosh(g\beta))\right)-\beta\frac{-2\eta g}{1+\eta^2}V-V\log(2\cosh(g\beta))\notag\\
    &=\beta\left[-gV\tanh(g\beta)+\frac{2\eta g}{1+\eta^2}V\right],
\end{align}
which agrees with the extractable work in Eq.~(\ref{work1}).

\vskip\baselineskip
\underline{$\rm(\hspace{.08em}ii\hspace{.08em})$ Case $(1+\eta)^2e^{-g\beta}\le (1-\eta)^2e^{g\beta}$ (low temperature or low measurement accuracy)}
\vskip\baselineskip

In this case, the feedback unitary operator $u_+$ is
\begin{align}
    u_+&=c\sigma^0,
\end{align}
where $|c|^2=1$. Under this feedback operation,
\begin{align}
    u_+m_{+}\rho^1_i m_{+}^{\dagger}u_+^{\dagger}=\frac{1}{4(1+\eta^2)\cosh(g\beta)}\left[(1+\eta)^2e^{-g\beta}\ket{\uparrow}\bra{\uparrow}+(1-\eta)^2e^{g\beta}\ket{\downarrow}\bra{\downarrow}\right].\label{123-4}
\end{align}
Therefore,
\begin{align}
\rho_f^1&:=u_+ m_+ u_1 \rho_i^1 u_1^{\dagger}m_+^{\dagger}u_+^{\dagger}+u_- m_- u_1 \rho_i^1 u_1^{\dagger}m_-^{\dagger}u_-^{\dagger}=\frac{1}{2\cosh(g \beta)}\left[e^{-g\beta}\ket{\uparrow}\bra{\uparrow}+e^{g\beta}\ket{\downarrow}\bra{\downarrow}\right]
\end{align}
and hence
\begin{align}
    {\rm Tr}[H_{f}^1\rho_{f}^1]=-g\tanh(g\beta),
\end{align}
where
\begin{align}
    H_f^1:=g\sigma^3.
\end{align}

By performing the same calculation as in Case $\rm(\hspace{.18em}i\hspace{.18em})$, we obtain the extractable work as
\begin{align}
    W_{\rm opt, loc}&=0.\label{work2}
\end{align}
Thus, no work can be extracted when the system temperature or the measurement accuracy is below a certain threshold.

Next, we compute the corresponding information quantities, namely, the mutual information, the Holevo information, and the nonequilibrium contribution. We also compute their sum, which is the right-hand side of Eq.~(\ref{eq:canonical-identity}), and check that it indeed agrees with the extractable work.

First, we compute
$\sum_{{\bm{k}}}P_{\bm{k}} S(\rho_{\bm{k}}^{\prime})$, where
\begin{align}
    P_{{\bm{k}}}:=\prod_{l=1}^{V}p_{k_l}
\end{align}
with
\begin{align}
    p_+&={\rm Tr}[m_{+}^{\dagger}m_{+}\rho_i^1]=\frac{(1+\eta)^2 e^{-g\beta}+(1-\eta)^2 e^{g\beta}}{4(1+\eta^2)\cosh(g\beta)},
\end{align}
\begin{align}
    p_-&=1-p_{+}=\frac{(1-\eta)^2 e^{-g\beta}+(1+\eta)^2 e^{g\beta}}{4(1+\eta^2)\cosh(g\beta)}.
\end{align}

Also,
\begin{align}
    S(\rho_{{\bm{k}}}^{\prime})=\sum_{l=1}^{V}S(\rho_{k_l}^1)
\end{align}
and
\begin{align}
    \sum_{{\bm{k}}}P_{\bm{k}} S(\rho_{\bm{k}}^{\prime})=V\left(\sum_{k_1=\pm}p_{k_1}S(\rho_{k_1}^1)\right).
\end{align}

From Eq.~(\ref{123-4}),
\begin{align}
    \rho_+^{1}&=\frac{u_+ m_+ u_1 \rho_i^1 u_1^{\dagger}m_+^{\dagger}u_+^{\dagger}}{p_+}=\frac{1}{4p_+(1+\eta^2)\cosh(g\beta)}\left[(1+\eta)^2e^{-g\beta}\ket{\uparrow}\bra{\uparrow}+(1-\eta)^2e^{g\beta}\ket{\downarrow}\bra{\downarrow}\right],
\end{align}
and from Eq.~(\ref{123-2}),
\begin{align}
    \rho_-^{1}&=\frac{u_- m_- u_1 \rho_i^1 u_1^{\dagger}m_-^{\dagger}u_-^{\dagger}}{p_-}=\frac{1}{4p_-(1+\eta^2)\cosh(g\beta)}\left[(1-\eta)^2e^{-g\beta}\ket{\uparrow}\bra{\uparrow}+(1+\eta)^2e^{g\beta}\ket{\downarrow}\bra{\downarrow}\right].
\end{align}

Hence,
\begin{align}
    &\sum_{{\bm{k}}}P_{\bm{k}} S(\rho_{\bm{k}}^{\prime})=V\left(\sum_{k_1=\pm}p_{k_1}S(\rho_{k_1}^1)\right)\notag\\
    &=V\left(-\frac{(1+\eta)^2e^{-g\beta}}{4(1+\eta^2)\cosh(g\beta)}\log\left[\frac{(1+\eta)^2e^{-g\beta}}{(1+\eta)^2e^{-g \beta}+(1-\eta)^2e^{g \beta}}\right]\right.\left.
    -\frac{(1-\eta)^2e^{-g\beta}}{4(1+\eta^2)\cosh(g\beta)}\log\left[\frac{(1-\eta)^2e^{-g\beta}}{(1-\eta)^2e^{-g \beta}+(1+\eta)^2e^{g \beta}}\right]\right.\notag\\
    &\left.
    -\frac{(1-\eta)^2e^{g\beta}}{4(1+\eta^2)\cosh(g\beta)}\log\left[\frac{(1-\eta)^2e^{g\beta}}{(1+\eta)^2e^{-g \beta}+(1-\eta)^2e^{g \beta}}\right]\right.\left.
    -\frac{(1+\eta)^2e^{g\beta}}{4(1+\eta^2)\cosh(g\beta)}\log\left[\frac{(1+\eta)^2e^{g\beta}}{(1-\eta)^2e^{-g \beta}+(1+\eta)^2e^{g \beta}}\right]
    \right).
\end{align}

Next, we compute $S(\rho_i^{\can})$. Since the von Neumann entropy of the initial state is the same as in Case $\rm(\hspace{.18em}i\hspace{.18em})$, we have
\begin{align}
    S(\rho_i^{\can})=V\left(-\frac{e^{-g \beta}}{2\cosh(g\beta)}\log\left[\frac{e^{-g \beta}}{2\cosh(g\beta)}\right]-\frac{e^{g \beta}}{2\cosh(g\beta)}\log\left[\frac{e^{g \beta}}{2\cosh(g\beta)}\right]\right).
\end{align}

Next, 
using Eqs.~(\ref{123-4}) and (\ref{123-2}), we obtain
\begin{align}
    &S(\rho_f)=VS\left[\frac{1}{4(1+\eta^2)\cosh(g\beta)}\left(2(1+\eta^2)e^{-g\beta}\ket{\uparrow}\bra{\uparrow}+2(1+\eta^2)e^{g\beta}\ket{\downarrow}\bra{\downarrow}\right)\right]\label{rho_f_can}\\
    &=V\left[-\frac{e^{-g\beta}}{2\cosh(g\beta)}\log\left(\frac{e^{-g\beta}}{2\cosh(g\beta)}\right)-\frac{e^{g\beta}}{2\cosh(g\beta)}\log\left(\frac{e^{g\beta}}{2\cosh(g\beta)}\right)\right],
\end{align}
which is equal to the von Neumann entropy $S(\rho_i^{\can})$ of the initial state:
\begin{align}
    S(\rho_f)=S(\rho_i^{\can}).
\end{align}

From the above results, the mutual information $I_{\rm QC}$ is
\begin{align}
    I_{\rm QC}&=S(\rho_i^{\can})-\sum_{{\bm{k}}}P_{\bm{k}} S(\rho_{\bm{k}}^{\prime})\notag\\
    &=V\left(-\frac{e^{-g \beta}}{2\cosh(g\beta)}\log\left[\frac{e^{-g \beta}}{2\cosh(g\beta)}\right]-\frac{e^{g \beta}}{2\cosh(g\beta)}\log\left[\frac{e^{g \beta}}{2\cosh(g\beta)}\right]\right)\notag\\
    &-V\left(-\frac{(1+\eta)^2e^{-g\beta}}{4(1+\eta^2)\cosh(g\beta)}\log\left[\frac{(1+\eta)^2e^{-g\beta}}{(1+\eta)^2e^{-g \beta}+(1-\eta)^2e^{g \beta}}\right]\right.\notag\\
    &\left.
    -\frac{(1-\eta)^2e^{-g\beta}}{4(1+\eta^2)\cosh(g\beta)}\log\left[\frac{(1-\eta)^2e^{-g\beta}}{(1-\eta)^2e^{-g \beta}+(1+\eta)^2e^{g \beta}}\right]\right.\notag\\
    &\left.
    -\frac{(1-\eta)^2e^{g\beta}}{4(1+\eta^2)\cosh(g\beta)}\log\left[\frac{(1-\eta)^2e^{g\beta}}{(1+\eta)^2e^{-g \beta}+(1-\eta)^2e^{g \beta}}\right]\right.\notag\\
        &\left.
    -\frac{(1+\eta)^2e^{g\beta}}{4(1+\eta^2)\cosh(g\beta)}\log\left[\frac{(1+\eta)^2e^{g\beta}}{(1-\eta)^2e^{-g \beta}+(1+\eta)^2e^{g \beta}}\right]
    \right),
\end{align}
which is the same as in Case $\rm(\hspace{.18em}i\hspace{.18em})$.

Next, we compute the Holevo information.
\begin{align}
    \chi&=S(\rho_f)-\sum_{{\bm{k}}}P_{\bm{k}} S(\rho_{\bm{k}}^{\prime})\notag\\
    &=V\left[-\frac{e^{-g\beta}}{2\cosh(g\beta)}\log\left(\frac{e^{-g\beta}}{2\cosh(g\beta)}\right)-\frac{e^{g\beta}}{2\cosh(g\beta)}\log\left(\frac{e^{g\beta}}{2\cosh(g\beta)}\right)\right]\notag\\
    &-V\left(-\frac{(1+\eta)^2e^{-g\beta}}{4(1+\eta^2)\cosh(g\beta)}\log\left[\frac{(1+\eta)^2e^{-g\beta}}{(1+\eta)^2e^{-g \beta}+(1-\eta)^2e^{g \beta}}\right]\right.\notag\\
    &\left.
    -\frac{(1-\eta)^2e^{-g\beta}}{4(1+\eta^2)\cosh(g\beta)}\log\left[\frac{(1-\eta)^2e^{-g\beta}}{(1-\eta)^2e^{-g \beta}+(1+\eta)^2e^{g \beta}}\right]\right.\notag\\
    &\left.
    -\frac{(1-\eta)^2e^{g\beta}}{4(1+\eta^2)\cosh(g\beta)}\log\left[\frac{(1-\eta)^2e^{g\beta}}{(1+\eta)^2e^{-g \beta}+(1-\eta)^2e^{g \beta}}\right]\right.\notag\\
        &\left.
    -\frac{(1+\eta)^2e^{g\beta}}{4(1+\eta^2)\cosh(g\beta)}\log\left[\frac{(1+\eta)^2e^{g\beta}}{(1-\eta)^2e^{-g \beta}+(1+\eta)^2e^{g \beta}}\right]
    \right).
\end{align}

Next, we compute the relative entropy.
Since $\rho_f^1=\rho_i^1$ in this case, $D(\rho_f^\can\Vert\rho_i^{\can})=0$:
\begin{align}
    D(\rho_f^\can\Vert\rho_i^{\can})&=-S(\rho_f)+\beta{\rm Tr}[H_f \rho_f]+\log Z\notag\\
    &=-V\left[-\frac{e^{-g\beta}}{2\cosh(g\beta)}\log\left(\frac{e^{-g\beta}}{2\cosh(g\beta)}\right)-\frac{e^{g\beta}}{2\cosh(g\beta)}\log\left(\frac{e^{g\beta}}{2\cosh(g\beta)}\right)\right]-\beta gV\tanh(g\beta)+V\log(2\cosh(g\beta))\notag\\
    &=0.
\end{align}

Finally, we compute the sum of the information quantities, i.e., the right-hand side of Eq.~(\ref{eq:canonical-identity}), and check that it indeed agrees with the extractable work.
\begin{align}
    &I_{\rm QC}-\chi-D(\rho_f^\can\Vert\rho_i^{\can})\notag\\
    &=V\left(-\frac{e^{-g \beta}}{2\cosh(g\beta)}\log\left[\frac{e^{-g \beta}}{2\cosh(g\beta)}\right]-\frac{e^{g \beta}}{2\cosh(g\beta)}\log\left[\frac{e^{g \beta}}{2\cosh(g\beta)}\right]\right)\notag+\beta gV\tanh(g\beta)-V\log(2\cosh(g\beta))\notag\\
    &=0,
\end{align}
which agrees with the extractable work in Eq.~(\ref{work2}).

\end{document}